\documentclass[twocolumn,prd,preprintnumbers,superscriptaddress,nofootinbib]{revtex4-2}

\usepackage{amsmath}
\usepackage{graphicx}
\usepackage{siunitx}
\usepackage{hyperref}
\usepackage[all]{hypcap}
\usepackage{booktabs}
\usepackage{multirow}
\usepackage{tabularx}
\usepackage{makecell}
\usepackage[dvipsnames]{xcolor}
\usepackage{listofitems}
\usepackage{fp}
\usepackage{longtable}

\DeclareSIUnit\ev{\electronvolt}
\DeclareSIUnit{\c}{\ensuremath{\mathit{c}}}

\setsepchar{\\/,/ }
\readlist\skonly{2.40,0.07,0.09\\0.450,0.059,0.026\\-1.885,0.865,1.180\\0.020,0.016,0.011\\1022.0616}
\readlist\skonlyi{2.40,0.05,0.33\\0.475,0.073,0.051\\-1.885,1.324,1.969\\0.010,0.021,0.008\\1027.2899}
\FPsub\skonlydmo{\skonlyi[5]}{\skonly[5]}
 
\readlist\skqfix{2.40,0.07,0.09\\0.450,0.057,0.026\\-1.745,0.760,1.252\\0.020,0.015,0.010\\1022.0565}
\readlist\skqfixi{2.40,0.06,0.12\\0.450,0.078,0.028\\-1.745,0.889,1.217\\0.010,0.016,0.007\\1027.7470}
\FPsub\skqfixdmo{\skqfixi[5]}{\skqfix[5]}

\readlist\skqfixp{0.88132525,0.0091135375\\0.41704625,0.019475135\\0.98516025,0.0032628229}
\FPsub\resA{1}{\skqfixp[1,1]}
\FPsub\resB{1}{\skqfixp[2,1]}
\FPsub\resC{1}{\skqfixp[3,1]}
\FPdiv{\skqfixclsA}{\skqfixp[1,2]}{\resA}
\FPdiv{\skqfixclsB}{\skqfixp[2,2]}{\resB}
\FPdiv{\skqfixclsC}{\skqfixp[3,2]}{\resC}
\readlist\skqfixcls{\skqfixclsA\\\skqfixclsB\\\skqfixclsC}
\FPeval\skqfixpref{100.*(1-\skqfixclsA)}

\DeclareRobustCommand{\sk}[1]{%
  \setsepchar{,}%
  \readlist\qarg{#1}%
  SK~\qarg[1]%
  \ifnum\listlen\qarg[]>1\relax--\qarg[2]\fi%
}
\newcommand{\dcp}{\ensuremath{\delta_{\text{CP}}}}
\newcommand{\sq}[2]{\ensuremath{\sin^{2}\theta_{#1#2}}}
\newcommand{\qq}[2]{\ensuremath{\theta_{#1#2}}}
\newcommand{\dms}[2]{\ensuremath{\Delta m^2_{#1#2}}}
\newcommand{\dmo}{\ensuremath{\Delta \chi^2_{\text{I.O.} - \text{N.O.}}}}
\newcommand{\cls}{\ensuremath{\text{CL}_s}}
\newcommand{\sklt}{6511.298}
\newcommand{\skfv}{27.1625}
\FPeval{\skexposure}{\sklt*\skfv/365.25}

\newcommand{\figref}[1]{Fig.~\ref{fig:#1}}
\newcommand{\Figref}[1]{Figure~\ref{fig:#1}}
\newcommand{\tabref}[1]{Tab.~\ref{tab:#1}}
\newcommand{\Tabref}[1]{Table~\ref{tab:#1}}
\newcommand{\eqnref}[1]{Eq.~(\ref{eq:#1})}
\newcommand{\Eqnref}[1]{Equation~(\ref{eq:#1})}
\newcommand{\secref}[1]{Sec.~\ref{sec:#1}}
\newcommand{\Secref}[1]{Section~\ref{sec:#1}}

\begin{document}

\title{Atmospheric neutrino oscillation analysis with neutron tagging and an expanded fiducial volume in Super-Kamiokande I--V}

\newcommand{\AFFicrr}{\affiliation{Kamioka Observatory, Institute for Cosmic Ray Research, University of Tokyo, Kamioka, Gifu 506-1205, Japan}}
\newcommand{\AFFkashiwa}{\affiliation{Research Center for Cosmic Neutrinos, Institute for Cosmic Ray Research, University of Tokyo, Kashiwa, Chiba 277-8582, Japan}}
\newcommand{\AFFipmu}{\affiliation{Kavli Institute for the Physics and
Mathematics of the Universe (WPI), The University of Tokyo Institutes for Advanced Study,
University of Tokyo, Kashiwa, Chiba 277-8583, Japan }}
\newcommand{\AFFmad}{\affiliation{Department of Theoretical Physics, University Autonoma Madrid, 28049 Madrid, Spain}}
\newcommand{\AFFubc}{\affiliation{Department of Physics and Astronomy, University of British Columbia, Vancouver, BC, V6T1Z4, Canada}}
\newcommand{\AFFbu}{\affiliation{Department of Physics, Boston University, Boston, MA 02215, USA}}
\newcommand{\AFFuci}{\affiliation{Department of Physics and Astronomy, University of California, Irvine, Irvine, CA 92697-4575, USA }}
\newcommand{\AFFcsu}{\affiliation{Department of Physics, California State University, Dominguez Hills, Carson, CA 90747, USA}}
\newcommand{\AFFcnm}{\affiliation{Institute for Universe and Elementary Particles, Chonnam National University, Gwangju 61186, Korea}}
\newcommand{\AFFduke}{\affiliation{Department of Physics, Duke University, Durham NC 27708, USA}}
\newcommand{\AFFgifu}{\affiliation{Department of Physics, Gifu University, Gifu, Gifu 501-1193, Japan}}
\newcommand{\AFFgist}{\affiliation{GIST College, Gwangju Institute of Science and Technology, Gwangju 500-712, Korea}}
\newcommand{\AFFuh}{\affiliation{Department of Physics and Astronomy, University of Hawaii, Honolulu, HI 96822, USA}}
\newcommand{\AFFicl}{\affiliation{Department of Physics, Imperial College London , London, SW7 2AZ, United Kingdom }}
\newcommand{\AFFkek}{\affiliation{High Energy Accelerator Research Organization (KEK), Tsukuba, Ibaraki 305-0801, Japan }}
\newcommand{\AFFkobe}{\affiliation{Department of Physics, Kobe University, Kobe, Hyogo 657-8501, Japan}}
\newcommand{\AFFkyoto}{\affiliation{Department of Physics, Kyoto University, Kyoto, Kyoto 606-8502, Japan}}
\newcommand{\AFFliv}{\affiliation{Department of Physics, University of Liverpool, Liverpool, L69 7ZE, United Kingdom}}
\newcommand{\AFFmiyagi}{\affiliation{Department of Physics, Miyagi University of Education, Sendai, Miyagi 980-0845, Japan}}
\newcommand{\AFFnagoya}{\affiliation{Institute for Space-Earth Environmental Research, Nagoya University, Nagoya, Aichi 464-8602, Japan}}
\newcommand{\AFFkmi}{\affiliation{Kobayashi-Maskawa Institute for the Origin of Particles and the Universe, Nagoya University, Nagoya, Aichi 464-8602, Japan}}
\newcommand{\AFFpol}{\affiliation{National Centre For Nuclear Research, 02-093 Warsaw, Poland}}
\newcommand{\AFFsuny}{\affiliation{Department of Physics and Astronomy, State University of New York at Stony Brook, NY 11794-3800, USA}}
\newcommand{\AFFokayama}{\affiliation{Department of Physics, Okayama University, Okayama, Okayama 700-8530, Japan }}
\newcommand{\AFFosaka}{\affiliation{Department of Physics, Osaka University, Toyonaka, Osaka 560-0043, Japan}}
\newcommand{\AFFox}{\affiliation{STFC Rutherford Appleton Laboratory, Harwell, Oxford, OX11 0QX, United Kingdom}}
\newcommand{\AFFqmul}{\affiliation{School of Physics and Astronomy, Queen Mary University of London, London, E1 4NS, United Kingdom}}
\newcommand{\AFFregina}{\affiliation{Department of Physics, University of Regina, 3737 Wascana Parkway, Regina, SK, S4SOA2, Canada}}
\newcommand{\AFFseoul}{\affiliation{Department of Physics, Seoul National University, Seoul 151-742, Korea}}
\newcommand{\AFFsheff}{\affiliation{Department of Physics and Astronomy, University of Sheffield, S3 7RH, Sheffield, United Kingdom}}
\newcommand{\AFFshizuokasc}{\affiliation{Department of Informatics in
Social Welfare, Shizuoka University of Welfare, Yaizu, Shizuoka, 425-8611, Japan}}
\newcommand{\AFFskk}{\affiliation{Department of Physics, Sungkyunkwan University, Suwon 440-746, Korea}}
\newcommand{\AFFtodai}{\affiliation{Department of Physics, University of Tokyo, Bunkyo, Tokyo 113-0033, Japan }}
\newcommand{\AFFtit}{\affiliation{Department of Physics,Tokyo Institute of Technology, Meguro, Tokyo 152-8551, Japan }}
\newcommand{\AFFtus}{\affiliation{Department of Physics, Faculty of Science and Technology, Tokyo University of Science, Noda, Chiba 278-8510, Japan }}
\newcommand{\AFFtoronto}{\affiliation{Department of Physics, University of Toronto, ON, M5S 1A7, Canada }}
\newcommand{\AFFtriumf}{\affiliation{TRIUMF, 4004 Wesbrook Mall, Vancouver, BC, V6T2A3, Canada }}
\newcommand{\AFFtokai}{\affiliation{Department of Physics, Tokai University, Hiratsuka, Kanagawa 259-1292, Japan}}
\newcommand{\AFFtsinghua}{\affiliation{Department of Engineering Physics, Tsinghua University, Beijing, 100084, China}}
\newcommand{\AFFynu}{\affiliation{Department of Physics, Yokohama National University, Yokohama, Kanagawa, 240-8501, Japan}}
\newcommand{\AFFllr}{\affiliation{Ecole Polytechnique, IN2P3-CNRS, Laboratoire Leprince-Ringuet, F-91120 Palaiseau, France }}
\newcommand{\AFFbari}{\affiliation{ Dipartimento Interuniversitario di Fisica, INFN Sezione di Bari and Universit\`a e Politecnico di Bari, I-70125, Bari, Italy}}
\newcommand{\AFFnapoli}{\affiliation{Dipartimento di Fisica, INFN Sezione di Napoli and Universit\`a di Napoli, I-80126, Napoli, Italy}}
\newcommand{\AFFroma}{\affiliation{INFN Sezione di Roma and Universit\`a di Roma ``La Sapienza'', I-00185, Roma, Italy}}
\newcommand{\AFFpadova}{\affiliation{Dipartimento di Fisica, INFN Sezione di Padova and Universit\`a di Padova, I-35131, Padova, Italy}}
\newcommand{\AFFkeio}{\affiliation{Department of Physics, Keio University, Yokohama, Kanagawa, 223-8522, Japan}}
\newcommand{\AFFwinnipeg}{\affiliation{Department of Physics, University of Winnipeg, MB R3J 3L8, Canada }}
\newcommand{\AFFkcl}{\affiliation{Department of Physics, King's College London, London, WC2R 2LS, UK }}
\newcommand{\AFFwarwick}{\affiliation{Department of Physics, University of Warwick, Coventry, CV4 7AL, UK }}
\newcommand{\AFFwu}{\affiliation{Faculty of Physics, University of Warsaw, Warsaw, 02-093, Poland }}
\newcommand{\AFFbcit}{\affiliation{Department of Physics, British Columbia Institute of Technology, Burnaby, BC, V5G 3H2, Canada }}
\newcommand{\AFFtohoku}{\affiliation{Department of Physics, Faculty of Science, Tohoku University, Sendai, Miyagi, 980-8578, Japan }}
\newcommand{\AFFicise}{\affiliation{Institute For Interdisciplinary Research in Science and Education, ICISE, Quy Nhon, 55121, Vietnam }}
\newcommand{\AFFilance}{\affiliation{ILANCE, CNRS - University of Tokyo International Research Laboratory, Kashiwa, Chiba 277-8582, Japan}}
\newcommand{\AFFibs}{\affiliation{Center for Underground Physics, Institute for Basic Science (IBS), Daejeon, 34126, Korea}}
\newcommand{\AFFglasgow}{\affiliation{School of Physics and Astronomy, University of Glasgow, Glasgow, Scotland, G12 8QQ, United Kingdom}}
\newcommand{\AFFoecu}{\affiliation{Media Communication Center, Osaka Electro-Communication University, Neyagawa, Osaka, 572-8530, Japan}}

\AFFicrr
\AFFkashiwa
\AFFmad
\AFFbu
\AFFbcit
\AFFuci
\AFFcsu
\AFFcnm
\AFFduke
\AFFllr
\AFFgifu
\AFFgist
\AFFglasgow
\AFFuh
\AFFibs
\AFFicise
\AFFicl
\AFFbari
\AFFnapoli
\AFFpadova
\AFFroma
\AFFilance
\AFFkeio
\AFFkek
\AFFkcl
\AFFkobe
\AFFkyoto
\AFFliv
\AFFmiyagi
\AFFnagoya
\AFFkmi
\AFFpol
\AFFsuny
\AFFokayama
\AFFoecu
\AFFox
\AFFseoul
\AFFsheff
\AFFshizuokasc
\AFFskk
\AFFtohoku
\AFFtokai
\AFFtodai
\AFFipmu
\AFFtit
\AFFtus
\AFFtriumf
\AFFtsinghua
\AFFwu
\AFFwarwick
\AFFwinnipeg
\AFFynu

\author{T.~Wester}
\AFFbu
\author{K.~Abe}
\AFFicrr
\AFFipmu
\author{C.~Bronner}
\AFFicrr
\author{Y.~Hayato}
\AFFicrr
\AFFipmu
\author{K.~Hiraide}
\AFFicrr
\AFFipmu
\author{K.~Hosokawa}
\AFFicrr
\author{K.~Ieki}
\author{M.~Ikeda}
\AFFicrr
\AFFipmu
\author{J.~Kameda}
\AFFicrr
\AFFipmu
\author{Y.~Kanemura}
\author{R.~Kaneshima}
\author{Y.~Kashiwagi}
\AFFicrr
% ...


\collaboration{The Super-Kamiokande Collaboration}
\noaffiliation

\date{\today}

\begin{abstract}
We present a measurement of neutrino oscillation parameters with the Super-Kamiokande detector using atmospheric neutrinos from the complete pure-water \sk{I,V} (April 1996--July 2020) data set, including events from an expanded fiducial volume. The data set corresponds to \num[round-mode=places,round-precision=1]{\sklt} live days and an exposure of \num[round-mode=places,round-precision=1]{\skexposure}~kiloton-years. Measurements of the neutrino oscillation parameters \dms{3}{2}, \sq{2}{3}, \sq{1}{3}, \dcp, and the preference for the neutrino mass ordering are presented with atmospheric neutrino data alone, and with constraints on \sq{1}{3}{} from reactor neutrino experiments. Our analysis including constraints on \sq{1}{3}{} favors the normal mass ordering at the \qty[round-mode=places,round-precision=1]{\skqfixpref}{\percent} level.
\end{abstract}

\maketitle

\section{Introduction}
\label{sec:intro}

Neutrino oscillations in the Pontecorvo-Maki-Nakagawa-Sakata (PMNS) paradigm are parameterized by three mixing angles, two squared-mass differences and a CP-violating phase~\cite{ponte1952,mns1962}. Experiments measuring neutrinos of different flavors, energies, and baselines have constrained many of the PMNS parameters with increasing levels of precision. However, the octant of the mixing angle \qq{2}{3}, the phase \dcp, and the sign of the larger of the two squared-mass differences, \dms{3}{2}, which determines the neutrino mass ordering, are all presently unknown. To date, the long-baseline accelerator neutrino experiments T2K~\cite{t2k_2021} and NOvA~\cite{nova_2022} have made the world's most precise measurements of \qq{2}{3}, \dms{3}{2}, and \dcp, but have yet to definitively resolve the remaining questions.

Atmospheric neutrinos are an independent and natural counterpart to accelerator neutrinos for studying neutrino oscillations. Neutrinos created in the earth's atmosphere span a range of energies and baselines that make their oscillations sensitive to the \qq{2}{3} mixing angle and the magnitude of the \dms{3}{2} squared-mass difference. Additionally, atmospheric neutrinos which pass near or through the dense core of the earth experience matter effects which alter their oscillation probabilities. An observation of these modified oscillation probabilities in either atmospheric neutrino or anti-neutrino data would provide important information towards resolving the neutrino mass ordering.

In this work, we analyze \num{6511.3} live days of atmospheric neutrino data from the Super-Kamiokande (SK) detector. This analysis improves upon the previous work~\cite{sk_atm_2018} in three major ways: We use the number of tagged neutrons to enhance the separation of neutrino events from anti-neutrino events, we enhance the efficiency of classifying multi-ring events using a boosted decision tree (BDT), and we add \qty{48}{\percent} exposure by analyzing events from an expanded fiducial volume and from \num{1186} additional live-days, including data collected after a major detector refurbishment in 2018. In addition to the atmospheric-only analysis, we present an analyses of SK data with an external constraint on the mixing angle \qq{1}{3} from the average measurements of the reactor neutrino experiments Daya Bay~\cite{dayabay_2018}, RENO~\cite{reno_2020} and Double-Chooz~\cite{doublechooz_2020}.

The paper is organized as follows: \Secref{intro} presents an overview of neutrino oscillation phenomenology relevant to atmospheric neutrino oscillations. \Secref{detector} provides a description of the Super-Kamiokande detector and its capabilities for reconstructing neutrino interactions. \Secref{simulation} describes the simulation used to model atmospheric neutrinos interactions at SK. \Secref{oa_skonly} describes the analysis methodology and presents the results of the analyses without external constraints and with constraints on \sq{1}{3}. We provide an interpretation and summary of the results in \Secref{interp}.

\subsection{\label{sec:oscillation}Neutrino Oscillations}

Neutrinos are produced as flavor eigenstates of the weak interaction, which may be treated as superpositions of mass eigenstates via the PMNS matrix:
\begin{equation}
\lvert \nu_{\alpha}\rangle = \sum_{i=1}^3 \mathbf{U}_{\alpha i}^* \lvert \nu_i \rangle,
\end{equation}

\noindent where $\alpha$ is a label for each lepton flavor, one of $e$, $\mu$, or $\tau$, and $\mathbf{U}_{\alpha i}$ is an element of the PMNS matrix. The PMNS matrix is parameterized by three mixing angles and a phase, and factorizes into three sub-matrices which describe rotations by each mixing angle from the neutrino mass basis into the neutrino flavor basis:
\begin{align}
\label{eq:pmns}
\textbf{U} & = 
\begin{pmatrix}
1 & 0 & 0 \\
0 & c_{23} & s_{23} \\
0 & -s_{23} & c_{23}
\end{pmatrix}
\begin{pmatrix}
c_{13} & 0 & s_{13}e^{-i\dcp} \\
0 & 1 & 0 \\
-s_{13}e^{i\dcp} & 0 & c_{13}
\end{pmatrix} \nonumber \\
& \hspace{2em} \times
\begin{pmatrix}
c_{12} & s_{12} & 0 \\
 -s_{12} & c_{12} &  0\\
0 & 0 & 1 \\
\end{pmatrix}.
\end{align}

\noindent In \eqnref{pmns}, sines and cosines of the mixing angles are written as $\cos\theta_{ij}\equiv c_{ij}$ and  $\sin\theta_{ij}\equiv s_{ij}$ respectively, and  the phase \dcp{} changes sign for the anti-neutrino case. The probability of a neutrino of one flavor $\lvert \nu_{\alpha} \rangle$ oscillating to a different flavor $\lvert \nu_{\beta}\rangle$ after some time $t$, or, equivalently, along a baseline $L$, is found by computing the amplitude $\lvert\langle \nu_{\beta} \rvert \nu_{\alpha} \rangle\rvert^2$. The probability is nonzero for the case $\alpha \neq \beta$ if the mass states have nonzero mass differences given by the signed quantity $\dms{i}{j} = m_i^2 - m_j^2$.

In the simplest case, neutrinos oscillate in a vacuum and oscillation probabilities may be computed by propagating neutrino states according to their vacuum Hamiltonian, written here in the mass basis:
\begin{equation}
H_{\text{Vacuum}} = 
\begin{pmatrix}
\frac{m_1^2}{2E} & 0 & 0 \\
0  & \frac{m_2^2}{2E} & 0  \\
0 & 0 & \frac{m_3^2}{2E}  \\
\end{pmatrix}.
\end{equation}

This leads to oscillation probabilities of the form:
\begin{align}
\label{eq:oscprob}
P_{\alpha\to\beta} =& \delta_{\alpha \beta} - 4 \sum_{i>j} \operatorname{Re}\left(U_{\alpha i}^*U_{\beta i} U_{\alpha j}U_{\beta j}^*\right)\sin^2 \Delta_{ij} \nonumber \\
& \pm 2 \sum_{i>j} \operatorname{Im}\left(U_{\alpha i}^*U_{\beta i} U_{\alpha j}U_{\beta j}^*\right)\sin 2\Delta_{ij},
\end{align}

\noindent where $\Delta_{ij} = 1.27 \dms{i}{j}L/E$. Here, \dms{i}{j} is expressed in units of \unit{\ev\squared}, $L$ is the oscillation baseline in kilometers, and $E$ is in the neutrino energy in \unit{\giga\ev}. Experiments have measured all mixing angles and squared mass differences to be significantly different from zero, while the value of the phase \dcp{} is still unknown\footnote{Recent results from T2K favor maximal CP violation, $\dcp\approx-\pi/2$~\cite{t2knature_2020}, while recent measurements from NOvA disfavor CP-violating scenarios~\cite{nova_2022}.}. In addition, solar neutrino oscillation experiments observe evidence for matter effects in the sun which imply the $\dms{2}{1}$ squared mass difference is positive, establishing an ordering for two of the neutrino masses, $m_2 > m_1$~\cite{sno_2001, sk_2001, wolfenstein_1978, ms_1985}. However, current experiments are consistent with either the \textit{normal} ordering, $m_3 \gg m_2, m_1$, or the \textit{inverted} ordering, $m_2, m_1 \gg m_3$. Consequently, the sign of the squared mass difference between $m_3$ and the next-most-massive neutrino, given by either \dms{3}{2} or \dms{3}{1}, is not known. We use the notation \dms{32}{,31} or simply $\Delta m^2$ for this squared mass difference where the ordering is not explicitly specified.

Numerically, \dms{32}{,31} has been measured to be approximately \num{30} times larger than \dms{2}{1}, such that $\dms{3}{2} \approx \dms{3}{1}$. The difference in magnitude between \dms{2}{1}{} and \dms{32}{,31}{} also implies that the terms in \eqnref{oscprob} containing one or the other squared mass differences dominate for different ranges of $L/E$. For long-baseline beam and atmospheric neutrinos, where neutrino baselines range from tens of kilometers to several thousand kilometers, and typical neutrino energies range from \unit{\mega\ev} to several \unit{\giga\ev}, the \dms{2}{1}{} terms are sub-dominant, leading to approximate flavor oscillation probabilities of the form
\begin{align}
\label{eq:osc_approx}
P(\nu_{\mu} \leftrightarrow \nu_{e}) & \approx \sin^2\theta_{23} \sin^2 2 \theta_{13} \sin^2\left(1.27 \frac{ \Delta m^2 L}{E}\right), \nonumber\\
P(\nu_{\mu} \to \nu_{\mu}) & \approx 1- 4\cos^2 \theta_{13} \sin^2 \theta_{23} (1-\cos^2 \theta_{13} \sin^2\theta_{23}) \nonumber \\
& \hspace{2em} \times \sin^2 \left(1.27 \frac{\Delta m^2 L }{E} \right), \nonumber \\ 
P(\nu_{e} \to \nu_e) & \approx 1-\sin^2 2\theta_{13} \sin^2\left(1.27\frac{\Delta m^2 L}{E}\right).
\end{align}

\noindent The approximate oscillation probabilities in \eqnref{osc_approx} are primarily functions of the mixing  angles and the absolute value of the squared-mass difference $\Delta m^2$. The phase \dcp, and the neutrino mass ordering, i.e., the sign of the squared-mass difference, are sub-leading effects which make them challenging experimental signatures. 

Neutrino oscillations in matter enhance the dependence of oscillation probabilities on the neutrino mass ordering. In matter, due to an increased forward scattering amplitude, electron-flavor neutrinos experience a larger potential relative to $\mu$ and $\tau$ flavors which modifies the vacuum Hamiltonian via an additional term,
\begin{equation}
H_{\text{Matter}} = 
H_{\text{Vacuum}}
+ \mathbf{U}^{\dagger}
\begin{pmatrix}
a & 0 & 0\\
0 & 0 & 0\\
0  & 0 & 0 \\
\end{pmatrix}
\mathbf{U},
\end{equation}

\noindent where $a=\pm \sqrt{2}G_F N_e$. Here, $G_F$ is the Fermi constant, $N_e$ is the electron density, and $\mathbf{U}$ is the PMNS matrix. The sign of $a$ is positive for neutrinos and negative for anti-neutrinos. Propagating the neutrino states according to the matter Hamiltonian leads to an effective squared-mass difference and mixing angle: 
\begin{align}
\label{eq:matter_mixing}
\Delta m^2_{\text{M}} &= \Delta m^2 \sqrt{ \sin^2 2 \theta_{13} + \left(\Gamma -\cos 2 \theta_{13}\right)^2}, \nonumber\\
\sin^2 2\theta_{13,{\text{M}}} &= \frac{\sin^22\theta_{13}}{\sin^2 2\theta_{13} + \left(\Gamma - \cos 2\theta_{13}\right)^2 },
\end{align}

\noindent where $\Gamma \equiv 2aE/\Delta m^2$. \Eqnref{matter_mixing} shows that the effective quantities depend on the sign of $\Delta m^2$. In particular, for neutrinos in the normal ordering, $\Gamma \approx \cos 2 \theta_{13}$ maximizes the effective mixing angle \sq{1}{3,\text{M}}. A maximum also occurs for anti-neutrinos in the inverted ordering. This maximum effective mixing angle predicts a resonant enhancement of muon-to-electron flavor conversions for either neutrinos or anti-neutrinos according to the neutrino mass ordering.

\subsection{Atmospheric Neutrinos}

\begin{table}
\caption{\label{tab:prem}Neutrino propagation layers and corresponding densities used for calculating neutrino oscillation probabilities in this analysis, based on a simplified PREM~\cite{prem1981}.}
\begin{ruledtabular}
\begin{tabular}{
l
S[table-format=4]
S[table-format=4]
S[table-format=2.1]
}
    Layer & {$R_{\text{Min.}}$ (\unit{\km})} & {$R_{\text{Max.}}$ (\unit{\km})} & {Density (\unit{\gram\per\cm\cubed})}\\
    \midrule
    Atmosphere & 6371 & {--} & 0\\
    Crust & 5701 & 6371 & 3.3 \\
    Mantle & 3480 & 5701& 5.0\\
    Outer core & 1220 & 3480 & 11.3\\
    Inner core & 0 & 1220 & 13.0\\
\end{tabular}
\end{ruledtabular}
\end{table}

\begin{figure*}
\includegraphics[width=0.93\linewidth]{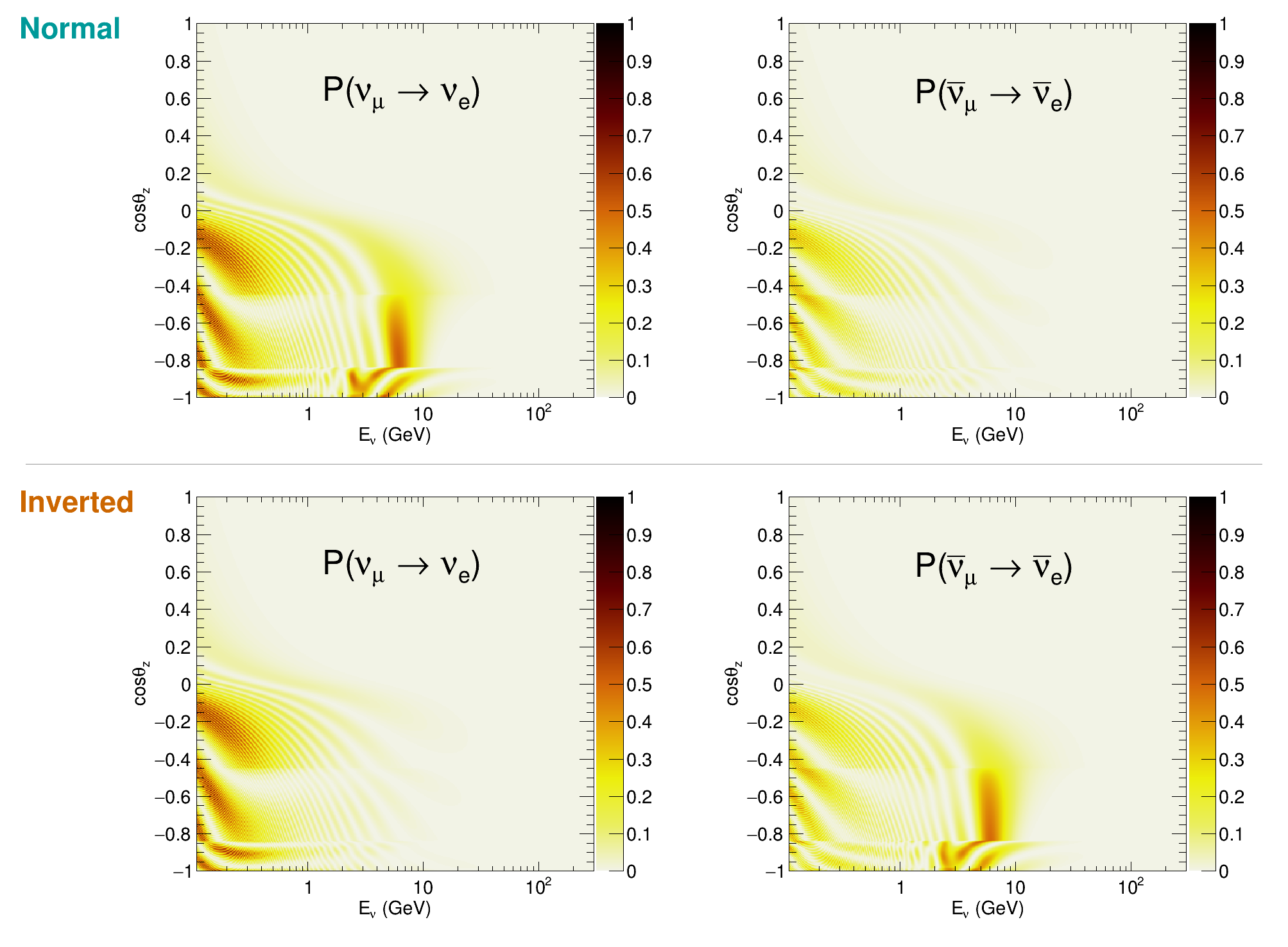}

\caption{\label{fig:oscillogram}Electron-to-muon flavor oscillation probabilities of atmospheric neutrinos as a function of cosine zenith angle and neutrino energy. The top row shows the probabilities for neutrinos and anti-neutrinos in the normal mass ordering scenario, and the bottom row shows the same probabilities for the inverted mass ordering scenario. The probabilities are calculated assuming $\sq{2}{3}=0.5$, $\sq{1}{3}=0.022$, $\sq{1}{2}=0.307$, $\lvert\dms{32}{,31}\lvert=\qty{2.4e-3}{\ev\squared}$, $\dms{2}{1}=\qty{7.53e-5}{\ev\squared}$, and $\dcp=-\pi/2$. The matter effect resonance is visible in the normal ordering for neutrinos (upper left) or the inverted ordering for anti-neutrinos (lower right) between \qtyrange{2}{10}{\giga\ev} and for $\cos \theta_z \lesssim -0.5$.}
\end{figure*}

Atmospheric neutrinos are produced when cosmic rays interact with nuclei in the earth's atmosphere. These interactions result in hadronic showers of primarily pions and kaons which decay into neutrinos. The atmospheric neutrino energy spectrum extends from a few \unit{\mega\ev} to several \unit{\tera\ev} and has an approximate flavor ratio in the few \unit{\giga\ev} range of $(\nu_{\mu} + \bar{\nu}_{\mu}) / (\nu_e + \bar{\nu}_e) \approx 2:1$. While present, tau neutrinos intrinsic to the atmospheric neutrino flux are suppressed by many orders of magnitude relative to electron- and muon-flavor neutrinos due to kinematic restrictions on their production.

The zenith angle $\theta_z$ describes atmospheric neutrino baselines. Neutrinos produced directly above a detector are downward going, $\theta_z=0$, and are produced at an average distance of \qty{15}{\km} above earth's surface. Neutrinos produced on the other side of the earth from a detector are upward-going, $\theta_z = \pi$, and travel an approximate distance of \qty{13000}{\km} through the earth. Oscillation signatures are most evident in upward-going atmospheric neutrinos due to the longer baselines.

A general atmospheric neutrino baseline begins at a production point in the atmosphere and passes through the earth before ending at a detector near the surface. We model the matter effects induced by passage through the earth assuming a simplified version of the preliminary reference Earth model (PREM)~\cite{prem1981}, where the earth is treated as a sphere with radius $R_{\text{Earth}}=\qty{6371}{\km}$ and contains concentric spherical shells of decreasing densities. \Tabref{prem} lists the earth layers and corresponding densities assumed in this work.

To compute neutrino oscillation amplitudes through layers of different matter densities, amplitudes along steps through matter of fixed densities are multiplied together~\cite{barger}. The general matrix form of the propagated mass eigenvectors $\mathbf{X}$ for neutrinos passing through a fixed matter density is
\begin{equation}
\mathbf{X} = \sum_k \left[\prod_{j\neq k} \frac{2 E H_{\text{Matter}} -M_j^2 \mathbf{I}}{M_k^2 -M_j^2} \right]\exp\left(-i \frac{M^2_k L}{2E } \right)
\end{equation}

\noindent where $M_i^2 /2E$ are the eigenvalues of $H_{\text{Matter}}$. This definition allows the neutrino probability along a baseline of changing matter density to be  written as
\begin{equation}
P(\nu_{\alpha} \to \nu_{\beta}) = \left|\left( \mathbf{U}  \prod_i \mathbf{X}(E, \rho_i, L_i) \mathbf{U}^{\dagger}\right)_{\alpha \beta} \right|^2,
\end{equation}

\noindent where $L_i$ and $\rho_i$ are the baseline and density of the $i^{\text{th}}$ step respectively. In the case of a spherically symmetric earth, as is assumed in this work, the neutrino oscillation baseline $L$ only depends on the zenith angle and production height and does not depend on the azimuth.

\Figref{oscillogram} shows the calculated oscillation probabilities for atmospheric neutrinos as a function of the cosine of the zenith angle and the neutrino energy $E_\nu$. The two sets of figures show the cases for muon- to electron-flavor neutrino and anti-neutrino oscillation probabilities in each mass ordering scenario, and assuming the current global average neutrino oscillation parameters~\cite{pdg2022}. The resonance in $\nu_{\mu}\to  \nu_e $ or $\bar{\nu}_{\mu} \to \bar{\nu}_e$ oscillations due to matter effects is exclusively visible in the normal and inverted scenarios respectively, and occurs at baselines of several thousand kilometers and neutrino energies around a few GeV. This resonance in atmospheric neutrinos is the experimental signature of the unknown neutrino mass ordering.

In addition to mass ordering sensitivity via earth matter effects, atmospheric neutrino oscillations also provide sensitivity to other oscillation parameters. Muon-to-tau flavor conversions provide sensitivity to $\lvert \dms{32}{,31}\lvert$ and \sq{2}{3}. While the tau neutrinos are often too low energy to produce CC interactions, the $P(\nu_{\mu}\to\nu_{\mu})$ survival probability manifests as a disappearance of upward-going muon neutrinos. Atmospheric neutrinos also provide modest sensitivity to the combined effects of \dcp{} and \sq{1}{3} through electron neutrino or anti-neutrino appearance for neutrinos of all energies.
\section{The Super-Kamiokande Detector}
\label{sec:detector}

Super-Kamiokande (SK) is a \num{50}~kiloton cylindrical water-Cherenkov detector located within the Kamioka mine in Gifu, Japan~\cite{skdet2003,skdetector_2014}. The detector consists of two optically separated regions: an inner detector (ID) which contains \num{32}~kilotons of water and is viewed by over \num{11000} inward-facing 20-inch photomultiplier tubes (PMTs), and a \qty{2}{\m} thick outer detector (OD) with over \num{1800} outward-facing 8-inch PMTs for vetoing cosmic backgrounds. To increase the light collected in the OD, the OD walls are covered with reflective Tyvek, and wavelength-shifting plates are mounted to the OD PMTs.

SK has been operational since its construction in 1996, and, until July 2020, has operated with pure water. During the this period, there were five distinct data-taking phases, \sk{I,V}, which had similar operating conditions with a few notable exceptions: At the end of the \sk{I} phase (1996--2001), an accident\footnote{A single PMT imploded during re-filling of the detector, creating a shockwave and chain reaction, destroying additional PMTs. Acrylic covers were mounted to all the PMTs to prevent similar accidents in the future.} resulted in the loss of roughly half of the experiment's PMTs. During the \sk{II} phase (2002--2005), the remaining PMTs were rearranged to provide uniform but reduced (\qty{19}{\percent}) photocoverage. New PMTs were installed to restore photocoverage to original conditions starting with the \sk{III} phase (2006--2008). In 2008, the experiment's electronics were upgraded~\cite{sk_nim_2009}, marking the start of the \sk{IV} phase (2008--2018). The \sk{IV} electronics upgrade extended the window of hit times recorded following neutrino-like events which improved the efficiency for detecting decay electrons and enabled the detection of the \qty{2.2}{\mega\ev} gamma emission following neutron captures on hydrogen~\cite{sk_ieee_2010}. 

During 2018, the SK detector was drained to conduct work in preparation of loading gadolinium sulfate, a compound with a high neutron capture cross section, into the detector's water. The work comprised of installing a new water circulation system capable of continuously purifying gadolinium sulfate, sealing the welding joints of the tank walls to repair and prevent leaks, and replacing several hundred PMTs that had failed during the \sk{III} and \sk{IV} phases. The subsequent \sk{V} (2019--2020) phase resumed data taking with pure water and reaffirmed the detector's stability and performance after the refurbishment work. In July 2020, gadolinium was dissolved into the detector's water for the first time~\cite{first_gd_loading}, marking the start of the \sk{Gd} phase. This work includes data from the pure water, \sk{I,V} phases only. Future analyses using data from the \sk{Gd} phase will feature enhanced neutron tagging efficiency due to the high neutron capture cross section of the gadolinium and its subsequent \qty{8}{\mega\ev} $\gamma$ cascade. The operating conditions of the SK phases are summarized in \tabref{sk_phases}.

\begin{table}
\caption{Super-Kamiokande data-taking phases. An electronics upgrade at the start of \sk{IV} enabled neutron tagging on hydrogen (H), utilized in the \sk{IV} and \sk{V} phases. During 2020, gadolinium (Gd) was added to the detector's water to increase the neutron-tagging efficiency. At the time of this writing, \sk{Gd} is ongoing and data from the \sk{Gd} phase are not included in this analysis.}
\begin{ruledtabular}
\begin{tabular}{@{}llS[table-number-alignment=center,table-format=4.1]cc@{}}
\multirow{2}*{Phase} & \multicolumn{1}{c}{\multirow{2}*{Dates}} & {Livetime}   & {Photo-} & Neutron\\ 
 &  & {(Days)}   & {coverage (\%)} & tagging\\ \midrule
    \sk{I} & 1996--2001 & 1489.2  & 40 & -- \\
    \sk{II} & 2002--2005 & 798.6  & 19 & -- \\
    \sk{III} & 2006--2008 & 518.1  & 40 & -- \\
    \sk{IV} & 2008--2018 & 3244.4  & 40 & H \\
    \sk{V} & 2019--2020 & 461.0 & 40  & H \\
    \midrule
    \sk{Gd} & 2020--Present  & {--}  & 40 & H+Gd \\
\end{tabular}
\end{ruledtabular}
\label{tab:sk_phases}
\end{table}

\begin{figure}
\includegraphics[width=0.99\columnwidth]{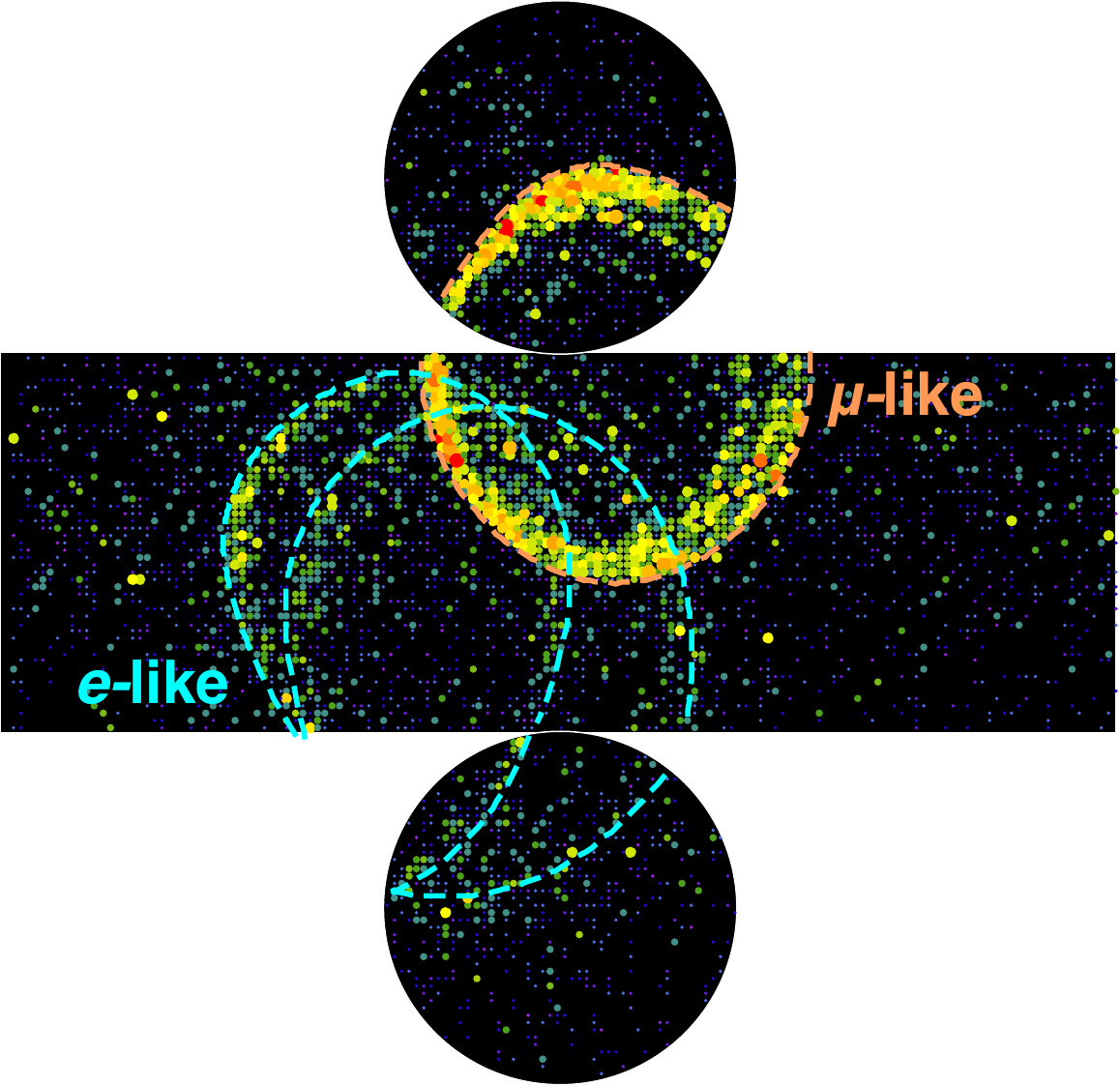}
\caption{\label{fig:event_display}Event display of a multi-ring atmospheric neutrino cadidate event in \sk{V} data. Hit ID PMTs are displayed on an unrolled view of the cylindrical detector, with the color and radius of each hit PMT corresponding to the detected charge. The reconstruction algorithm \textsc{APFit} identifies three Cherenkov rings, indicated by dashed outlines: one bright $\mu$-like ring, with $p_{\mu}\approx\qty{1010}{\mega\ev\per \c}$, and two fainter $e$-like rings, each with $p_e \approx \qty{320}{\mega\ev\per \c}$.}
\end{figure}

The SK detector observes Cherenkov light from charged particles with sufficient momentum produced following neutrino interactions. The light projected onto the PMT-lined walls of the detector forms ring patterns of hit PMTs. The ring patterns are reconstructed using the \textsc{APFit}~\cite{shiozawa_1999} algorithm: The timing information of hits establishes an event vertex, and a fit considering the spatial distribution and observed charges of hit PMTs determines the particle's direction, momentum, and particle type. \textsc{APFit} separates rings into $e$-like and $\mu$-like: Electrons and photons tend to scatter and produce electromagnetic showers, resulting in many overlapping rings which appear as a single ring with blurred edges. Heavier particles such as muons and charged pions do not create showers and therefore have sharp ring edges. Higher-momentum particles produce more light, so the charge contained within the ring provides an estimate of the particle's momentum. \Figref{event_display} shows an example of the Cherenkov light patterns observed in SK following a neutrino candidate interaction, and their fitted properties. The event contains multiple ring patterns, each corresponding to a different particle.

\begin{figure*}
\includegraphics[width=0.95\linewidth]{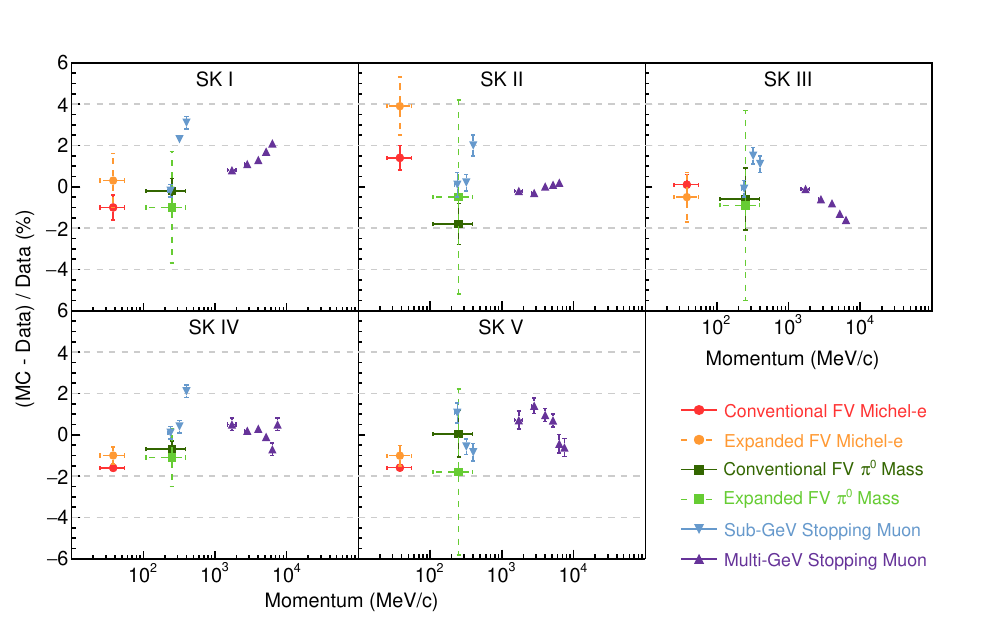}

\vspace{-0.6em}

\caption{\label{fig:escale}Performance of energy reconstruction for various sources across all SK periods. Solid points indicate measurements made within the conventional fiducial volume (vertex is greater than \qty{200}{\cm} from the detector walls) while dashed points correspond to measurements in the additional fiducial volume (vertex is between \qty{100}{\cm} and \qty{200}{\cm} from the detector walls). Michel-$e$ refers to the stopping muon decay electron spectrum calibration.}
\end{figure*}

In addition to Cherenkov rings, SK identifies electrons from muon decays. Decay electrons are found by scanning for time-clustered hits above background following a primary neutrino interaction trigger. A hit-time-based fitter estimates the decay electron vertex for each candidate hit cluster, and candidates are accepted if there are \num{50} or more hits within a \qty{50}{\ns} time window. The overall decay electron tagging efficiency is estimated to  be \qty{96}{\percent} for $\mu^+$ and \qty{80}{\percent} for $\mu^-$ in the \sk{IV} and \sk{V} periods. The reduced efficiency for $\mu^-$ is due to $\mu^-$ capture in the water, in which no decay electron is produced.

Neutrons in the SK detector are captured on hydrogen, producing deuterium in an excited state. The decay of the excited deuterium produces a \qty{2.2}{\mega\ev} $\gamma$ which results in a few time-coincident and spatially clustered PMT hits. These $\gamma$ emissions from neutron captures are identified using a two-step process~\cite{neutron2013}: In the first step, a sliding \qty{10}{\ns} hit-time window finds candidate neutron captures from clusters of \numrange{7}{50} hits with fewer than \num{200} hits in a surrounding \qty{200}{\ns} window. In the second step, hits within each candidate window are used to form variables which quantify the isotropy, likelihood of single-vertex origin, and the time spread of the hits. A neural network classifies candidates as either signal or background based on these variables. The average neutron tagging efficiency for neutron capture on hydrogen is \qty{26}{\percent} in the \sk{IV} and \sk{V} periods.

In the SK detector, the charge of particles, and therefore neutrino and anti-neutrino interactions, cannot be differentiated on an event-by-event basis. However, statistical separation is possible. For example, in the process $\bar{\nu}_{\mu} + p \to p + \mu^+ + \pi^-$, in which an anti-neutrino interacts with a proton $p$, the outgoing negatively-charged pion is more likely to be captured by an $\phantom{}^{16}\text{O}$ nucleus before decaying than is a positively-charged pion produced in the equivalent $\nu_{\mu}$ interaction. Captured pions do not produce decay electrons, so requiring one or more decay electrons preferentially selects more neutrino than anti-neutrino events for this process. The statistical separation can be further improved by also considering the number of neutrons, which will be described in \secref{detector:selection:neutron}.

\subsection{Calibration}
\label{sec:detector:calib}

\begin{figure}
\includegraphics[width=\columnwidth]{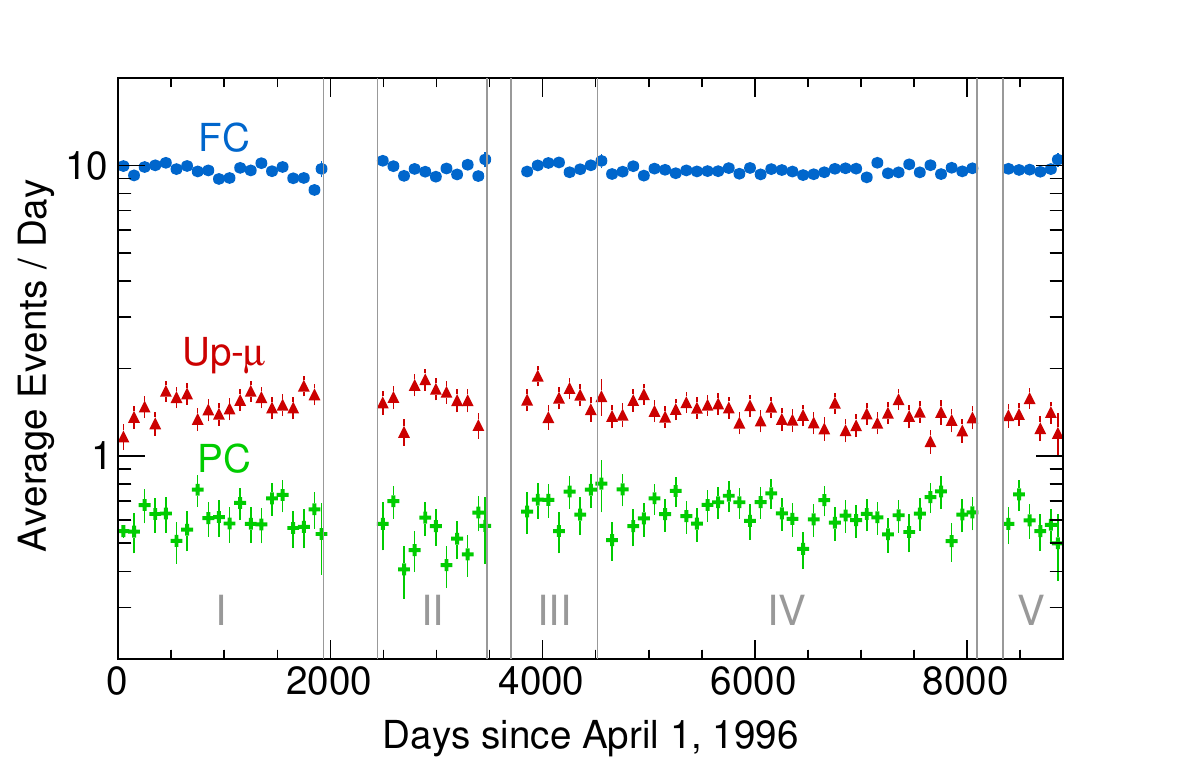}%

\vspace{-1em}

\caption{\label{fig:event_rate}Average event rates for the primary SK neutrino samples, fully contained (FC), partially contained (PC) and upward-going muons (Up-$\mu$). Error bars are statistical. The five pure-water data taking phases, \sk{I,V}, are labelled. The FC event rates includes events in the additional FV volume region, see \secref{detector:selection:exp_fv}.}
\end{figure}

Calibration ensures an accurate and consistent response of the detector to particle interactions. Calibration studies based on the detector geometry, PMT responses, and properties of the SK water are documented in Ref.~\cite{skdetector_2014}.

We assess \textsc{APFit}'s energy determination using calibration sources at multiple energies which span as much of the atmospheric neutrino energy spectrum as possible. Cosmic ray muons which stop within the detector and produce a decay electron provide a way to measure photoproduction as a function of the muon's track length. The track length of these muons is determined by their ID entrance point and decay electron vertex. The expected momentum of each muon assuming minimum-ionization along the track length may be compared to the fitted Cherenkov ring momentum. The energy spectrum of decay electrons from these muons also provides a low-energy calibration source. We use the momenta of the two $e$-like rings from neutral pion decay, $\pi^0\to\gamma \gamma$, which add to form an invariant mass distribution of neutral pion events, as a third calibration source. \Figref{escale} shows the difference between data and Monte Carlo (MC) for these calibration sources during each SK phase. In this analysis, the decay electron and $\pi^0$ mass measurements were also performed separately using events with vertices greater than $\qty{200}{\cm}$ (conventional fiducial volume) and between \qtyrange{100}{200}{\cm} (additional fiducial volume, see \secref{detector:selection:exp_fv}) from the detector walls.

\subsection{Neutrino Sample Selection}
\label{sec:detector:selection}

Neutrino events at SK are broadly categorized as fully-contained (FC), partially-contained (PC) or upward-going muons (Up-$\mu$). FC and PC events have a reconstructed event vertex within the ID and are differentiated based on the number of hits detected in the OD: FC events have minimal OD activity, while PC events have OD activity following the primary event trigger. Up-$\mu$ events are neutrino interactions within the rock below the SK tank or in the OD water which produce muons travelling upward. \Figref{event_rate} shows the average number of FC, PC, and Up-$\mu$ events observed per day during the \sk{I,V} phases.

We separate neutrino candidate interactions from each category into analysis samples to enhance the different oscillation signals present in the atmospheric neutrino data. The data taken during the \sk{IV} and \sk{V} periods, \num{3705}~days, or \qty{57}{\percent} of the total \sk{I,V} exposure, uses the observed number of neutron captures on hydrogen as an additional classification handle to enhance the purity of neutrino and anti-neutrino samples. 

\subsubsection{Analysis Samples}
\label{sec:detector:selection:standard}

\begin{figure*}
\includegraphics[width=1.0\textwidth]{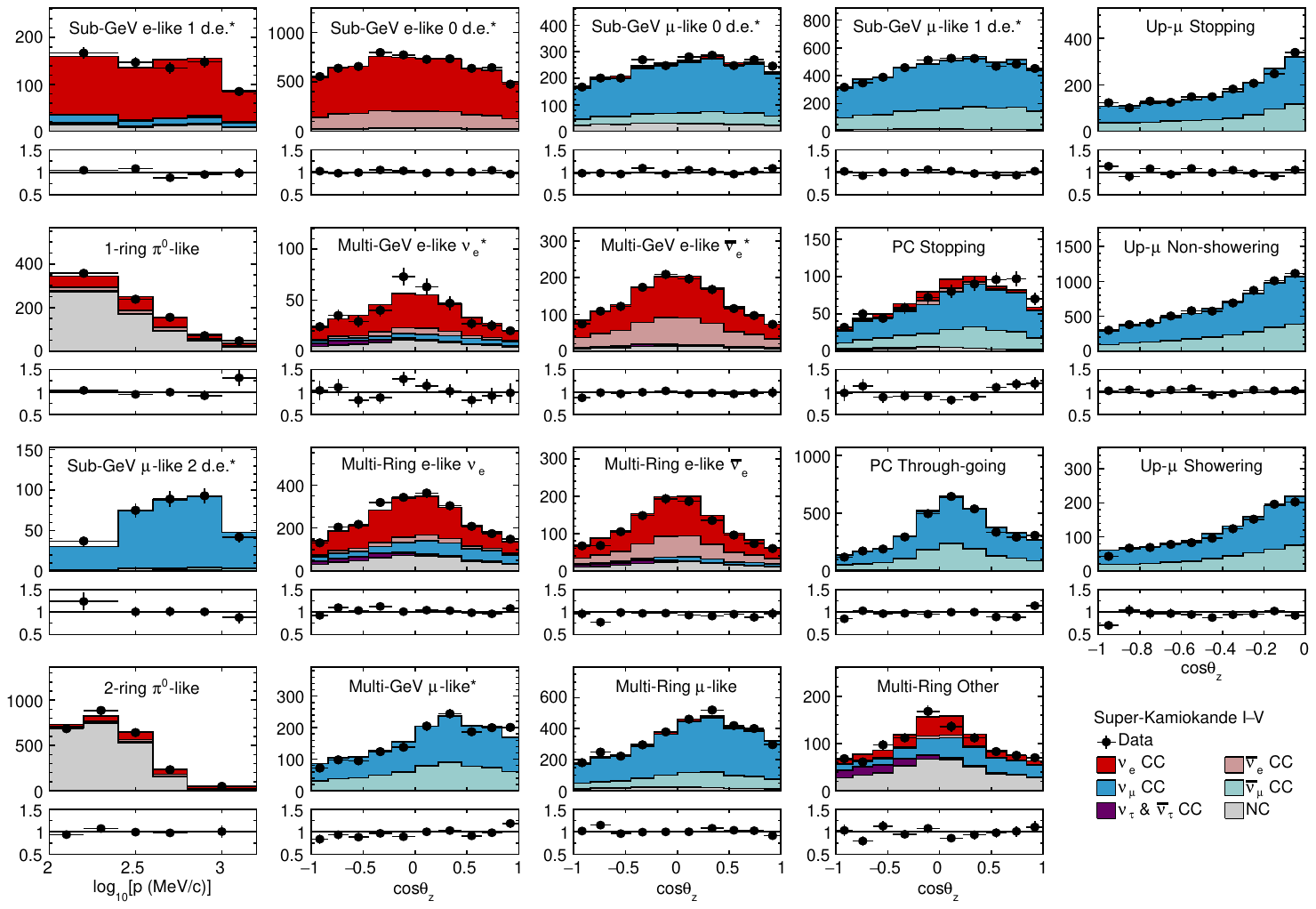}%

\vspace{-1em}

\caption{\label{fig:zenith_ratio_standard}Zenith angle or momentum distributions for the 19 analysis samples without neutron tagging. The first column shows the momentum bins for samples which are not also binned by zenith angle, while the second through fifth columns show the 1D zenith angle distributions for samples which use 2D momentum-and-zenith-angle binning. The MC distributions are shown at the best-fit point in the normal ordering, cf.\ \tabref{osc_grid}, with best-fit systematic pulls applied. Different colors in the MC histograms correspond to the true neutrino flavors present in each sample. Panels beneath each distribution show the data-MC ratio, and all error bars are statistical. FC (non-PC and non-Up-$\mu$) sub-GeV and multi-GeV single-ring samples, marked with an asterisk (*), contain events from only \sk{I,III}, while the corresponding events from \sk{IV,V} are separated into samples using tagged neutron information, shown in \figref{zenith_ratio_hybrid}.}
\end{figure*}

Fully contained events span the energy range of \qty{100}{\mega \ev} to \qty{100}{\giga \ev} and are a mixture of charged-current (CC) and neutral current (NC) neutrino interactions of all flavors. Because the normal and inverted neutrino mass ordering scenarios predict an enhancement to either the number of $\nu_{\mu}\to \nu_e$ or $\bar{\nu}_{\mu}\to \bar{\nu}_e$ events, and SK cannot distinguish the sign of neutrino interactions on an event-by-event basis, the FC sample definitions are designed to increase the statistical purity of $\nu_e$ and $\bar{\nu}_e$ events.

Fully contained events with a single Cherenkov ring are first separated by the ring's particle identification (PID) score, either $e$-like or $\mu$-like. Events are further divided based on the visible energy $E_{\text{vis.}}$ into sub-GeV, $E_{\text{vis.}}<\qty{1330}{\mega\ev}$, and multi-GeV, $E_{\text{vis.}}>\qty{1330}{\mega\ev}$. Next, events are separated by the number of decay electrons. For sub-GeV events, we use the number of decay electrons to separate events by likely interaction processes. Samples enhanced with quasi-elastic interactions are formed by requiring no decay electrons for $e$-like events and either zero or one decay electron for $\mu$-like events. Sub-GeV $e$-like and $\mu$-like events with one or more, or two or more decay electrons, respectively, are separated into additional samples enhanced in interaction processes which produce a pion below Cherenkov threshold. For multi-GeV events, the number of decay electrons, either zero, or one or more, is used to separate $e$-like events into anti-neutrino- and neutrino-enhanced samples. Fully contained single ring events in the \sk{IV} and \sk{V} phases have been revised in this analysis to additionally incorporate the number of tagged neutrons, and are discussed in \secref{detector:selection:neutron}.

Multi-ring events can contain mixtures of $e$-like and $\mu$-like rings, making the neutrino flavor ambiguous. However, multi-ring events also provide extra information which is useful for separating $\nu_e$ events from $\bar{\nu}_e$ events. For multi-GeV multi-ring events, we use a boosted decision tree (BDT) to classify these events as $\nu_e$-like, $\bar{\nu}_e$-like, $\mu$-like, or ``other.'' The ``other'' sample primarily selects NC events. More details on the BDT are presented in \secref{detector:selection:bdt}.

Sub-GeV multi-ring events are not included in the analysis due to their poor direction resolution and consequently minimal sensitivity to oscillation effects, with two exceptions: First, multi-ring events with $E_{\text{vis.}}>\qty{600}{\mega\ev}$ where the most energetic ring is $\mu$-like and has a reconstructed momentum of at least \qty{600}{\mega\ev\per \c} are included in the multi-ring $\mu$-like sample. Second, FC NC interactions which produce a $\pi^0$ are an oscillation-insensitive background to the other samples, but are included in the analysis to constrain NC interactions. These neutral current $\pi^0$ events are identified from sub-GeV events using a dedicated fitter which assumes there are two rings present, regardless of the number of reconstructed rings. Events are classified as $\pi^0$-like based on the likelihood that the two fitted rings originate from a $\pi^0$ decay. Events which are classified as $\pi^0$-like are separated into two samples based on the number of reconstructed rings without the two-ring assumption, either one or two.

\begin{figure*}
\includegraphics[width=1.0\textwidth]{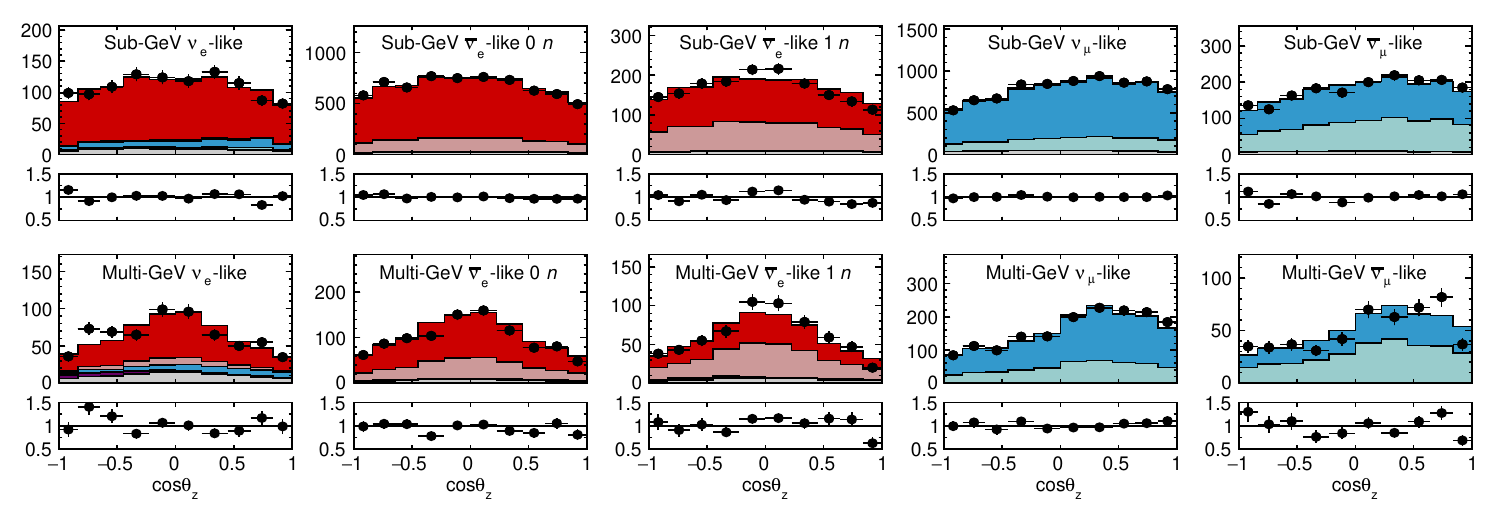}%

\vspace{-1em}

\caption{Zenith angle distributions for the \sk{IV} \& \sk{V} FC sub-GeV and multi-GeV single-ring samples using the number of tagged neutrons. The MC configuration and the meaning of the colors are identical to \figref{zenith_ratio_standard}.}
\label{fig:zenith_ratio_hybrid}
\end{figure*}

Partially contained events have typical energies between \qty{1}{\giga\ev} to \qty{1}{\tera\ev}, and are nearly all $\nu_{\mu}$ CC interactions, as the muon produced in the interaction often exits the ID. The muon momentum in PC events can only be estimated using the portion of the track within the detector. Partially contained events which exit the ID and stop within the OD, determined by comparing the amount of light in the OD to simulated PC events, are classified as stopping, while PC events that completely exit the detector are classified as through-going.

Upward-going muon events are the highest-energy events, up to $\sim\qty{10}{\tera\ev}$, observed at SK and are classified into three samples: ``stopping'' if the muon stops within the ID, otherwise ``showering'' or ``non-showering,'' depending on whether or not the exiting muon's charge deposition is consistent with radiative losses. The earth shields the Up-$\mu$ sample from cosmic ray backgrounds from below the horizon. However, upward-scattered muons from downward-going cosmic rays are an irreducible background near the horizon. We estimate the background rate in each Up-$\mu$ sample at the horizon by comparing the cosmic muon rate at the azimuthal directions with the smallest and largest mountain overburdens. The estimated number of background events is subtracted from the number of Up-$\mu$ data events with reconstructed directions near the horizon. The estimated Up-$\mu$ background rate is approximately $\numrange[range-units=single,range-phrase = - ]{2}{3}\,\unit{\percent}$.

The event classification outlined in this section corresponds to \num{19} distinct analysis samples. There are eight FC CC-enhanced single ring samples which are used for data collected during the \sk{I,III} phases, while the remaining \num{11} FC multi-ring, NC $\pi^0$, PC, and Up-$\mu$ samples are used for data taken during all phases. The data and MC counts for the \num{19} standard analysis samples are presented in \figref{zenith_ratio_standard} as a function of reconstructed lepton momentum or zenith angle.

\subsubsection{SK IV--V Neutron-Tagged Samples}
\label{sec:detector:selection:neutron}

This analysis modifies the FC single-ring event selection during the \sk{IV} and \sk{V} phases based on the number of observed neutron captures on hydrogen. The modification is motivated by the greater average neutron production in anti-neutrino interactions relative to neutrino interactions: Additional neutrons are expected in anti-neutrino events from proton-to-neutron ($p\to n$) conversions in CC processes, $\bar{\nu}_l + p\to l^{+} + n + X$, where $l$ denotes lepton flavor, and also in both CC and NC deep inelastic scattering (DIS) processes due to the larger fraction of energy transferred to the recoiling hadronic system. We incorporated neutron information into the analysis sample definitions to create 10 additional samples, five for sub-GeV events and five for multi-GeV events, as follows:

\begin{itemize}
\item \textit{FC single-ring $e$-like}: Both sub-GeV and multi-GeV events are divided into three samples: Events with one or more decay electron and any number of neutrons are classified as $\nu_e$-like. Events with no decay electrons are considered $\bar{\nu}_e$-like, and are further separated into two samples based on whether or not there is at least one tagged neutron.
\item \textit{FC single-ring $\mathbf{\mu}$-like}: Both sub-GeV and multi-GeV events are divided into two sub-samples: Events with exactly one decay electron and one or more tagged neutrons are considered $\bar{\nu}_{\mu}$-like, otherwise they are considered $\nu_{\mu}$-like.
\end{itemize}

\begin{table*}
\caption{Monte Carlo CC and NC purities by sample and data event counts used in this analysis. Purities and MC counts are shown with oscillation probabilities applied and without the effects of systematic pulls. The ``$\nu_{\tau}$ CC'' column shows the purity of both $\nu_{\tau}$ and $\bar{\nu}_{\tau}$ CC events. The Up-$\mu$ data counts are shown after background subtraction.}
\begin{ruledtabular}
\begin{tabular}{@{}lcc
*{6}{S[table-format=1.3]}
S[table-format=4.1]
S[table-format=4.1,round-mode=places,round-precision=0,table-number-alignment=right,table-alignment-mode=none]
@{}
}
\multirow{2}{*}{Sample} & 
\multirow{2}{*}{Energy bins} &
\multirow{2}{*}{$\cos\theta_{z}$ bins} & 
\multicolumn{6}{c}{MC purity} & \multicolumn{2}{c}{Events} \\
\cmidrule(lr){4-9} \cmidrule(l){10-11}
& & &
{$\nu_e$ CC} &
{$\bar{\nu}_e$ CC} &
{$\nu_{\mu}$ CC} &
{$\bar{\nu}_{\mu}$ CC} &
{$\nu_{\tau}$ CC} &
{NC} &
{MC} &
{Data} \\
\midrule
\multicolumn{11}{l}{\textit{Fully contained (FC), single ring, Sub-GeV}}\\
\noalign{\smallskip}
\multicolumn{11}{l}{SK I-III}\\
\multicolumn{11}{l}{$e$-like}\\
\hspace{0.5em}0 decay-$e$       &$5$ $e^{\pm}$ momentum &$10$ in $[-1,1]$       &     0.733     &     0.226     &     0.003     &     0.001     &
 0.000  &     0.036     &    6409.1     &    6647.0\\
\hspace{0.5em}1 decay-$e$       &$5$ $e^{\pm}$ momentum &single bin     &     0.796     &     0.016     &     0.086     &     0.020     &     0.001 &
     0.081      &     612.0     &     682.0\\
\noalign{\smallskip}
\multicolumn{11}{l}{$\mu$-like}\\
\hspace{0.5em}0 decay-$e$       &$5$ $\mu^{\pm}$ momentum       &$10$ in $[-1,1]$       &     0.027     &     0.008     &     0.704     &     0.149 &
     0.001      &     0.112     &    2153.9     &    2419.0\\
\hspace{0.5em}1 decay-$e$       &$5$ $\mu^{\pm}$ momentum       &$10$ in $[-1,1]$       &     0.001     &     0.000     &     0.677     &     0.291 &
     0.000      &     0.030     &    4241.4     &    4476.0\\
\hspace{0.5em}2 decay-$e$       &$5$ $\mu^{\pm}$ momentum       &single bin     &     0.001     &     0.000     &     0.948     &     0.029     &
 0.001  &     0.022     &     330.7     &     336.0\\
\noalign{\smallskip}
\multicolumn{11}{l}{SK IV-V}\\
\hspace{0.5em}$\nu_{e}$-like    &$5$ $e^{\pm}$ momentum &$10$ in $[-1,1]$       &     0.794     &     0.016     &     0.090     &     0.024     &
 0.001  &     0.074     &     943.7     &    1093.0\\
\hspace{0.5em}$\bar{\nu}_{e}$-like 0 $n$ &$5$ $e^{\pm}$ momentum &$10$ in $[-1,1]$       &     0.789     &     0.175     &     0.003     &
 0.001  &     0.000     &     0.031     &    5961.8     &    6669.0\\
\hspace{0.5em}$\bar{\nu}_{e}$-like 1 $n$      &$5$ $e^{\pm}$ momentum &$10$ in $[-1,1]$       &     0.582     &     0.367     &     0.003     &     0.002 &
     0.001      &     0.044     &    2266.6     &    1668.0\\
\hspace{0.5em}$\nu_{\mu}$-like  &$5$ $\mu^{\pm}$ momentum       &$10$ in $[-1,1]$       &     0.011     &     0.003     &     0.755     &     0.173 &
     0.000      &     0.057     &    6596.0     &    7879.0\\
\hspace{0.5em}$\bar{\nu}_{\mu}$-like    &$5$ $\mu^{\pm}$ momentum       &$10$ in $[-1,1]$       &     0.001     &     0.000     &     0.533     &
 0.417  &     0.001     &     0.049     &    2150.4     &    1793.0\\
\noalign{\smallskip}
\multicolumn{11}{l}{\textit{Fully contained (FC), single ring, Multi-GeV}}\\
\noalign{\smallskip}
\multicolumn{11}{l}{SK I-III}\\
\hspace{0.5em}$\nu_{e}$-like    &$4$ $e^{\pm}$ momentum &$10$ in $[-1,1]$       &     0.568     &     0.086     &     0.102     &     0.014     &
 0.039  &     0.190     &     359.6     &     383.0\\
\hspace{0.5em}$\bar{\nu}_{e}$-like      &$4$ $e^{\pm}$ momentum &$10$ in $[-1,1]$       &     0.556     &     0.341     &     0.013     &     0.002 &
     0.012      &     0.075     &    1359.8     &    1339.0\\
\hspace{0.5em}$\nu_{\mu}$-like  &$2$ $\mu^{\pm}$ momentum       &$10$ in $[-1,1]$       &     0.002     &     0.001     &     0.621     &     0.371 &
     0.003      &     0.002     &    1588.5     &    1564.0\\
\noalign{\smallskip}
\multicolumn{11}{l}{SK IV-V}\\
\hspace{0.5em}$\nu_{e}$-like    &$4$ $e^{\pm}$ momentum &$10$ in $[-1,1]$       &     0.607     &     0.087     &     0.098     &     0.015     &
 0.033  &     0.159     &     584.1     &     643.0\\
\hspace{0.5em}$\bar{\nu}_{e}$-like 0 $n$ &$4$ $e^{\pm}$ momentum &$10$ in $[-1,1]$       &     0.637     &     0.287     &     0.008     &
 0.002  &     0.007     &     0.058     &     866.4     &     986.0\\
\hspace{0.5em}$\bar{\nu}_{e}$-like 1 $n$    &$4$ $e^{\pm}$ momentum &$10$ in $[-1,1]$       &     0.435     &     0.460     &     0.009     &     0.002 &
     0.015      &     0.079     &     736.1     &     616.0\\
\hspace{0.5em}$\nu_{\mu}$-like  &$2$ $\mu^{\pm}$ momentum       &$10$ in $[-1,1]$       &     0.002     &     0.001     &     0.695     &     0.297 &     0.003      &     0.001     &    1464.0     &    1619.0\\
\hspace{0.5em}$\bar{\nu}_{\mu}$-like    &$2$ $\mu^{\pm}$ momentum       &$10$ in $[-1,1]$       &     0.001     &     0.000     &     0.446     &
 0.549  &     0.004     &     0.002     &     593.1     &     503.0\\
\midrule
\multicolumn{11}{l}{SK I-V common samples}\\
\noalign{\smallskip}
\multicolumn{11}{l}{\textit{Fully contained (FC) Sub-GeV NC $\pi^0$-like}}\\
\hspace{0.5em}Single-ring       &$5$ $e^{\pm}$ momentum &single bin     &     0.219     &     0.064     &     0.018     &     0.002     &     0.001 &     0.696      &     748.0     &     868.0\\
\hspace{0.5em}Two-ring          &$5$ $\pi^{0}$ momentum &single bin     &     0.096     &     0.028     &     0.015     &     0.001     &     0.000 &     0.860      &    2095.3     &    2494.0\\
\noalign{\smallskip}
\multicolumn{11}{l}{\textit{Fully contained (FC) Multi-GeV, multi-ring}}\\
\hspace{0.5em}$\nu_{e}$-like    &3 visible energy       &$10$ in $[-1,1]$       &     0.495     &     0.066     &     0.175     &     0.014     &
 0.035  &     0.215     &    2149.7     &    2411.0\\
\hspace{0.5em}$\bar{\nu}_{e}$-like      &3 visible energy       &$10$ in $[-1,1]$       &     0.519     &     0.260     &     0.052     &     0.007 &     0.025      &     0.138     &    1210.0     &    1131.0\\
\hspace{0.5em}$\mu$-like        &4 visible energy       &$10$ in $[-1,1]$       &     0.028     &     0.003     &     0.713     &     0.200     &
 0.006  &     0.050     &    3257.7     &    3427.0\\
\hspace{0.5em}Other             &4 visible energy       &$10$ in $[-1,1]$       &     0.203     &     0.023     &     0.257     &     0.014     &
 0.074  &     0.429     &     837.9     &     982.0\\
\noalign{\smallskip}
\multicolumn{11}{l}{\textit{Partially-contained (PC)}}\\
\hspace{0.5em}Stopping          &2 visible energy       &$10$ in $[-1,1]$       &     0.089     &     0.034     &     0.559     &     0.262     &
 0.011  &     0.045     &     641.6     &     689.0\\
\hspace{0.5em}Through-going     &4 visible energy       &$10$ in $[-1,1]$       &     0.006     &     0.002     &     0.638     &     0.341     &
 0.007  &     0.006     &    3310.2     &    3397.0\\
\noalign{\smallskip}
\multicolumn{11}{l}{\textit{Upward-going muons (Up-$\mu$)}}\\
\hspace{0.5em}Stopping          &3 visible energy       &$10$ in $[-1,0]$       &     0.008     &     0.003     &     0.646     &     0.340     &
 0.000  &     0.003     &    1574.3     &   {1753.8}\\
\hspace{0.5em}Non-showering     &single bin             &$10$ in $[-1,0]$       &     0.002     &     0.001     &     0.662     &     0.334     &
 0.000  &     0.001     &    5315.8     &    {6423.9}\\
\hspace{0.5em}Showering         &single bin             &$10$ in $[-1,0]$       &     0.001     &     0.000     &     0.671     &     0.327     &
 0.000  &     0.001     &    1051.4     &    {1110.6}\\
\end{tabular}
\end{ruledtabular}
\label{tab:purity}
\end{table*}

More details about the neutron tagging algorithm and the event selection modifications may be found in Refs.~\cite{neutron2013,pablo_2017}. \Figref{zenith_ratio_hybrid} shows the data and MC comparisons for the \sk{IV-V} neutron tagged samples. Compared to the equivalent anti-neutrino samples shown in \figref{zenith_ratio_standard}, the \sk{IV,V} $\bar{\nu}_e$  samples requiring at least one tagged neutron have a higher purity of true $\bar{\nu}_e$ MC events, summarized in \tabref{purity}.

\subsubsection{Multi-Ring Boosted Decision Tree}
\label{sec:detector:selection:bdt}

\begin{figure*}
\includegraphics[width=1.0\textwidth]{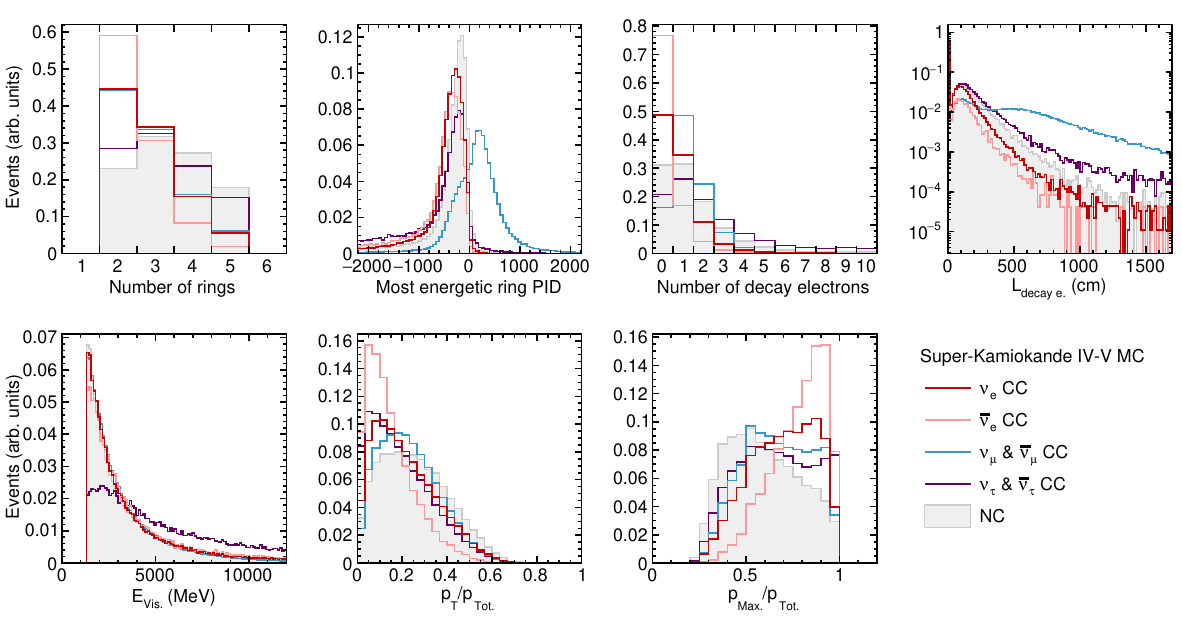}

\vspace{-1em}

\caption{\label{fig:mrbdt}Monte-Carlo distributions of input variables for the FC multi-GeV, multi-ring BDT. Distributions are area-normalized. $\text{L}_{\text{Decay e.}}$ is the maximum distance of any decay electron from the event's primary vertex. $p_{\text{T}}$ and $p_{Max.}$ are the transverse momentum and the momentum of the most energetic ring, respectively, and $p_{\text{Tot.}}$ is the sum of the momenta from all reconstructed rings.}
\end{figure*}

A previous version of this analysis~\cite{wendell_2010} classified multi-ring events into $e$-like or $\mu$-like using a likelihood function based on seven input variables: the number of rings, the number of decay electrons, a continuous PID variable of the most energetic ring present in the event, the fraction of energy carried by the most energetic ring, the transverse momentum of the most energetic ring, the maximum distance of any decay electrons from the event vertex, and the visible energy. The selection was performed in two steps, first to separate $e$-like events from $\mu$-like events, then to further separate $e$-like events into $\nu_e$-like and $\bar{\nu}_e$-like events. An additional ``other'' category was also present to separate NC events from CC events.

This analysis replaces the likelihood-based selection with a boosted decision tree (BDT) selection for multi-ring events, utilizing the same set of variables as in the likelihood-based classification~\cite{matsumoto_2020}. Labeled true multi-ring MC events are used to train the BDT classifier. Additionally, we optimize the BDT for sensitivity to the neutrino mass ordering by weighting these training events: Adjusting the weights of CC $\nu_e$ and $\bar{\nu}_e$ training events relative to CC $\nu_{\mu}$ and ``other'' events changes the expected signal purities obtained from the BDT selection, which in turn affects the expected mass ordering sensitivity. We find that the optimized BDT improves the efficiency for selecting CC events while maintaining an equivalent wrong-flavor contamination rate as the likelihood selection. The increased efficiency results in a higher proportion of correctly-classified CC $\nu_e$-like, $\bar{\nu}_e$-like, and $\nu_{\mu}$-like multi-ring events and a decreased proportion of ``other'' events relative to the previous analysis.

\begin{figure}
\includegraphics[width=1.0\linewidth]{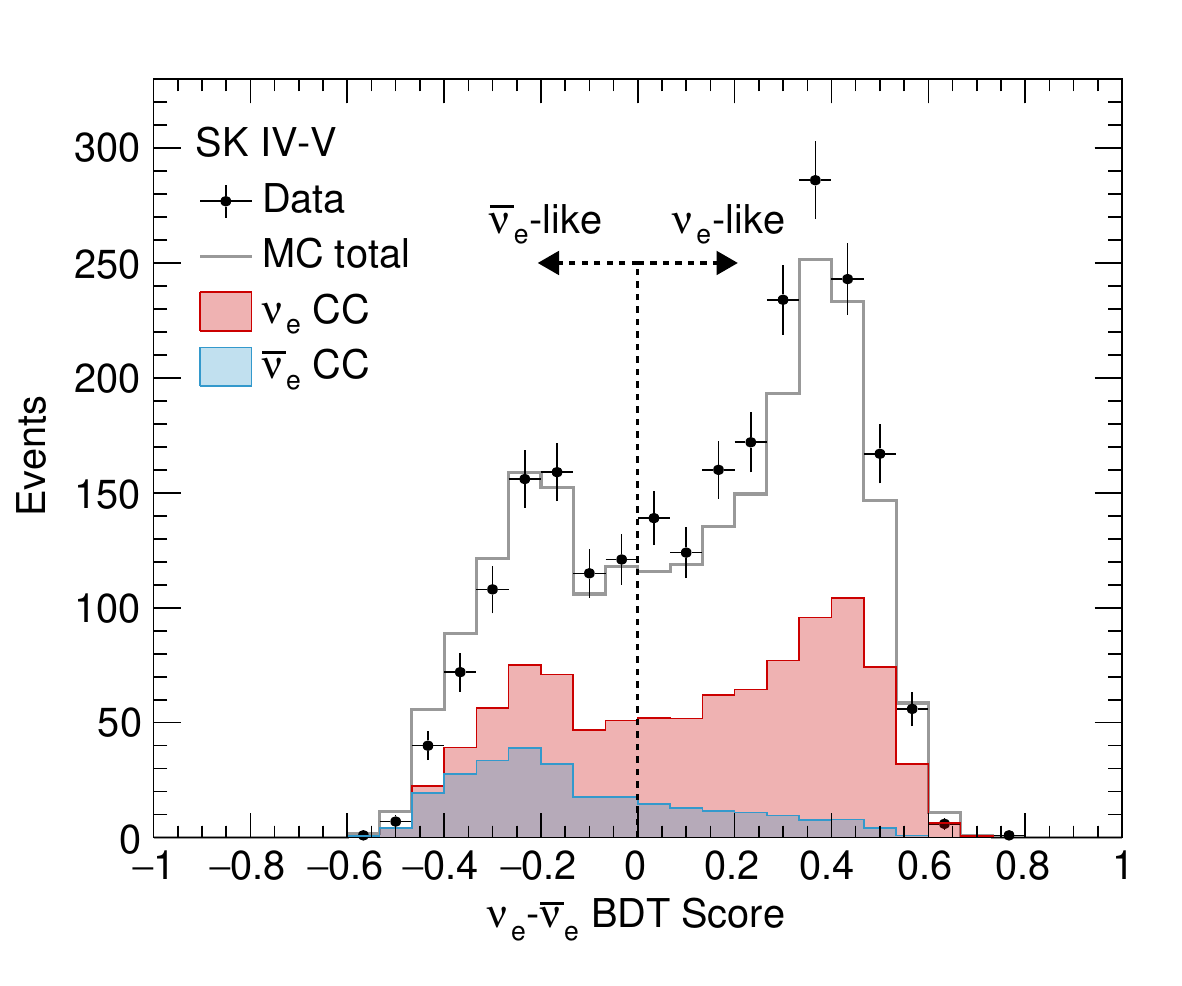}

\vspace{-1.3em}

\caption{\label{fig:bdt_separation}Data and MC distributions of the multi-ring BDT classifier scores for multi-ring $\nu_e$ and $\bar{\nu}_e$ events. The BDT makes a final classification decision for each event based on the class with the  highest score: Negative values correspond to $\bar{\nu}_e$-like and  positive values correspond to $\nu_e$-like. Distributions are shown for the \sk{IV,V} phases, including events in the expanded fiducial volume region. MC events are shown with oscillations applied, and without the effect of any fitted systematic uncertainty parameters. The ``MC total'' distribution includes contributions from true $\nu_{\mu}$, $\nu_{\tau}$, and NC events.}
\end{figure}

\Figref{mrbdt} shows the distributions of the BDT input variables from MC events. The largest differences between electron- and muon- flavor neutrino interactions are in the PID of the most energetic ring, the number of decay electrons, and the maximum distance travelled by decay electrons, $L_{\text{Decay} e}$. The difference in $L_{\text{Decay} e}$ between $\mu$-like and $e$-like events results from muons produced in $\nu_{\mu}$ CC interactions having higher momenta and traveling further on average than charged pions which are potentially produced in multi-ring interactions of any flavor. Electron neutrino and anti-neutrino interactions differ in the number of rings, and in the fractional and transverse momentum of the most energetic ring. These differences are a consequence of the different angular and momentum dependencies between neutrino and anti-neutrino cross sections. \Figref{bdt_separation} demonstrates the $\nu_{e}$ and $\bar{\nu}_e$ separation performance of the BDT in the \sk{IV,V} phases. Neutron information from the \sk{IV,V} phases was also considered as an input variable to the BDT, but was found to only result in a marginal improvement to $\nu_e$-$\bar{\nu}_e$ separation while introducing a large amount of systematic uncertainty.

\begin{figure*}
\includegraphics[width=1.0\textwidth]{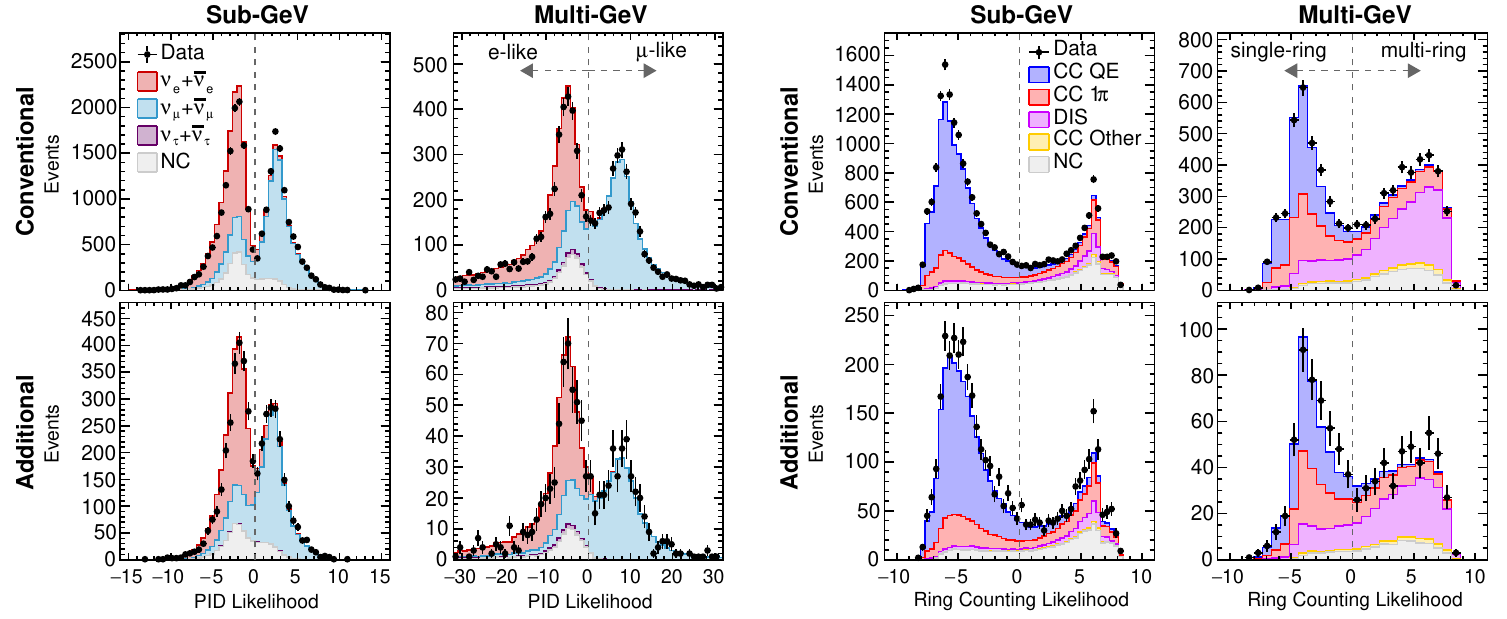}

\vspace{-1em}

\caption{\label{fig:pid_rc_likelihoods}\sk{IV,V} data and MC distributions of the likelihood variables used to classify events into the different analysis samples. The top row shows events in the conventional fiducial volume region, while the bottom row shows events in the additional fiducial volume region. The first and third columns show sub-GeV events and the second and fourth columns show multi-GeV events. The left four panels show the PID likelihood for the most energetic ring in each event: Negative values indicate the ring is $e$-like while positive values indicate the ring is $\mu$-like. The right four panels show the ring-counting likelihood: Negative values indicate the event has a single-ring, while positive values indicate the event has multiple rings. In these panels, the CC QE histograms include both 1p1h and 2p2h processes, and CC Other refers to single, non-pion hadron production. All MC distributions are calculated assuming neutrino oscillations and without the effect of any systematic uncertainties.}
\end{figure*}

\subsubsection{Expanded Fiducial Volume}
\label{sec:detector:selection:exp_fv}

Previous SK analyses used a \num{22.5}~kiloton fiducial volume for FC events defined by requiring the event vertex be at least \qty{200}{\cm} away from the ID walls (conventional fiducial volume). This analysis includes events in the \num{4.7}~kiloton region \qtyrange[range-units = single]{100}{200}{\cm} from the ID walls (additional fiducial volume) for all SK phases, representing a \qty{20}{\percent} increase in exposure. To include events in the additional fiducial volume region, both the reconstruction algorithms and background estimations required re-evaluation~\cite{takenaka_2020}. In 2018, the pre-tabulated expected charge distributions, which are the basis of the ring counting and PID likelihoods used in \textsc{APFit}, were re-computed as a function of a particle's distance to the nearest detector wall. These new charge tables encode the reduced number of hits and increased angular dependence expected from events near the detector walls. Adopting the updated charge tables reduced the $e$-$\mu$ mis-identification of single ring events in the additional region by up to \qty{35}{\percent}.

Cosmic muons with vertices mis-reconstructed in the fiducial volume also become more prevalent closer to the detector walls. To estimate the size of these backgrounds, events in the additional fiducial volume region were eye-scanned by experts. A slightly higher background rate of \qty{0.5}{\percent}, compared to \qty{0.1}{\percent} in the conventional fiducial volume, was found. The increased background rate in the additional fiducial volume is acceptable, although the background rate rises quickly for events with vertices $<\qty{100}{\cm}$, prohibiting further expansion.

We note that Ref.~\cite{takenaka_2020} included events from the additional fiducial volume region in the context of a proton decay search which was primarily concerned with sub-GeV atmospheric neutrino backgrounds. In contrast, this oscillation analysis includes both sub-GeV and multi-GeV atmospheric neutrino events. Additional studies to those presented in Ref.~\cite{takenaka_2020} were needed to confirm the performance of the reconstruction algorithms for multi-GeV events in the additional fiducial volume region. We studied the reconstructed momentum and direction biases and resolutions using MC events, and confirmed that these quantities were similar between events in the conentional and additional fiducial volume regions. We also studied the data versus MC agreement between the ring PID and ring counting likelihoods for events in the additional fiducial volume region, as shown in \figref{pid_rc_likelihoods}. The figure compares the data and MC distributions for sub-GeV and multi-GeV events in both fiducial volume regions. The level of agreement in the likelihood distributions using the updated reconstruction algorithms is found to be equivalent between events in the conventional and expanded fiducial volume regions at both sub-GeV and multi-GeV energies.

This analysis is the first SK atmospheric neutrino oscillation analysis to include FC events from the additional fiducial volume region, resulting in a total fiducial volume of \num[round-mode=places,round-precision=1]{\skfv}~kiloton for all SK phases. The total exposure for FC events in this analysis is \num[round-mode=places,round-precision=1]{\skexposure}~kiloton-years, a \qty{48}{\percent} increase over the previous published analysis.

\section{Simulation}
\label{sec:simulation}

We produce simulated Monte Carlo (MC) events to provide a prediction of atmospheric neutrino data. The SK MC consists of simulated neutrino interactions generated according to the flux model of Honda et al.~\cite{honda_2011} and the cross section models of the \textsc{neut} simulation software~\cite{neut,neut2021}. \textsc{neut} also propagates intermediate particles within the nuclear medium, and produces the final-state particles which exit the nucleus. These final-state particles are stepped through a \textsc{geant3}-based~\cite{geant_1994} simulation of the SK detector which models particle propagation in the detector water, Cherenkov radiation emission, and the detection of Cherenkov radiation by PMTs~\cite{skdetector_2014}. The SK detector simulation also implements a data-driven model of PMT electronics, dark noise, and disabled PMTs.

\subsection{Neutrino Interaction Model}
\label{sec:simulation:xsec}

\begin{figure}
\centering
\includegraphics[width=1.0\linewidth]{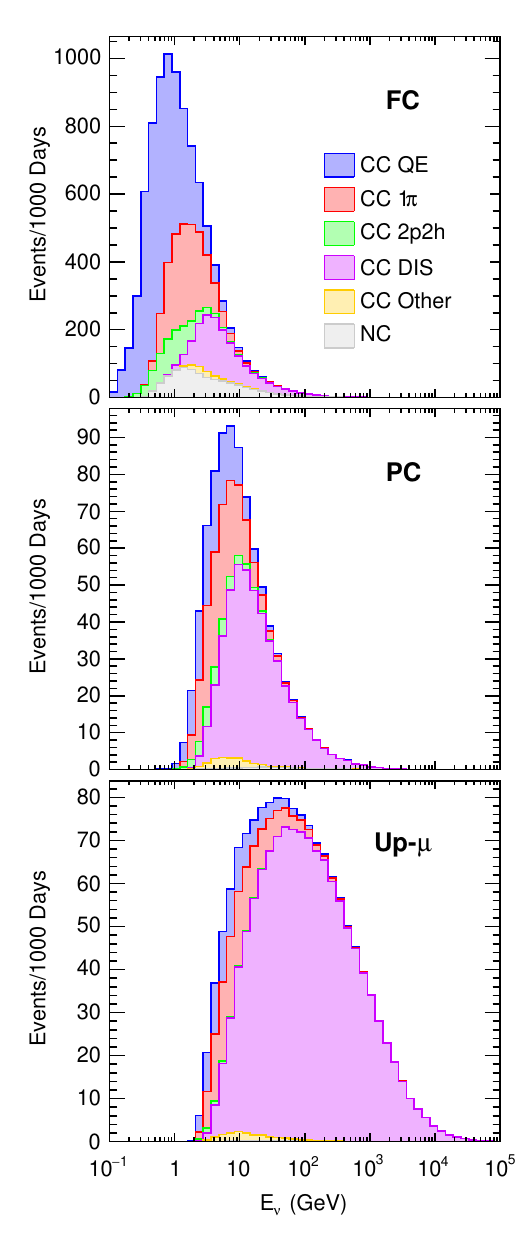}

\vspace{-1em}

\caption{Simulated neutrino energy spectra by interaction modes for the  different SK event categories: fully contained (FC), partially contained (PC) and upward-going  muon (Up-$\mu$). The simulated event rates are for an oscillated MC, with oscillation parameters set to the Particle Data Group best-fit values~\cite{pdg2022}, except for \dcp{} which  is set to $0$. ``CC Other'' modes refer to the production of single, non-pion hadrons.}
\label{fig:int_modes}
\end{figure}

Because atmospheric neutrinos span several orders of magnitude in energy, multiple interaction processes are relevant across the different SK neutrino samples. Fully contained events, which are the lowest-energy atmospheric neutrinos in this analysis, have leading contributions from charged-current quasi-elastic (CCQE) and single-pion production processes. Partially contained and Up-$\mu$ events consist of higher-energy neutrinos where deep inelastic scattering (DIS) becomes dominant. The contribution  to the event rate from each cross section process is shown in \figref{int_modes}.

This analysis uses \textsc{neut} version 5.4.0 \cite{neut2021} for the nominal neutrino interaction models, which is an update from version 5.3.6 used in the previous analysis. Notable changes in this version include the replacement of the nominal one-particle-one-hole (1p1h) model, responsible for the CCQE and NCQE processes, from the Smith and Moniz relativistic Fermi gas (RFG) model~\cite{smith_moniz_1972} to the local Fermi gas (LFG) model by Nieves et al.~\cite{nieves2011}. The nominal value of the axial mass parameter $M_A^{\text{QE}}$ was decreased from \qty{1.21}{\giga\ev\per \c\squared} to \qty{1.05}{\giga\ev\per \c \squared}. This change results in an overall decrease of CCQE events in the nominal MC by $\sim\qty{20}{\percent}$, although the previous $M_A^{\text{QE}}$ value is still within the $1\sigma$ uncertainty, see \secref{oa_skonly:syst}. The LFG model also includes a correction to the cross  section at low four-momentum ($q^2$) transfers due to weak charge screening calculated via the random phase approximation (RPA) technique~\cite{nieves_rpa}. The BBBA05~\cite{bbba05} vector form factors are used in the nominal 1p1h model, which is unchanged from the previous analysis.

\textsc{neut} implements the Rein-Sehgal resonant single-pion production model~\cite{rein_sehgal_1981}, which is unchanged from the previous analysis. The coherent pion production cross section calculation, which was previously based on the Rein-Sehgal model, has been updated to use the Berger-Sehgal model~\cite{berger_sehgal_2007} for neutrino events with $E_{\nu}<\qty{10}{\giga\ev}$. This was done to improve the agreement in total cross section and angular distribution of the outgoing pion with recent scattering experiments~\cite{martins2016charged}.

Deep inelastic scattering cross sections in \textsc{neut} are calculated using the GRV98 parton distribution functions (PDFs)~\cite{grv98}, with corrections to the low-$q^2$ regime from Bodek \& Yang~\cite{bodek_yang}. \textsc{neut} simulates the production of multiple hadrons using two separate models, selected based on the invariant mass of the hadronic system, $W$. For $W<\qty{2}{\giga\ev}$, a custom multi-pion generator is used, while for $W>\qty{2}{\giga\ev}$, \textsc{neut} uses \textsc{pythia} v5.72~\cite{pythia572}. The DIS hadron production models are only used to produce final states with multiple hadrons to avoid overlap with the single-hadron models.

Final state interactions (FSIs) are processes which modify particles exiting the nucleus due to intra-nuclear effects. Secondary interactions (SIs) are equivalent processes which occur in the detector medium instead of within the nucleus. \textsc{neut} implements four FSI+SI processes: quasi-elastic scattering, charge exchange, pion absorption, and hadron production. Six parameters adjust the probabilities of these processes. \textsc{neut} 5.4.0 updated the default values of these parameters~\cite{duet2017}, which increased the pion absorption probability by \qty{27}{\percent}, and decreased the probability of charge exchange for pions with momentum $<\qty{400}{\mega\ev\per \c}$ by \qty{30}{\percent} compared to version 5.3.6.

The cross section for the two-particle-two-hole (2p2h) process, in which no pions are produced but a pair of nucleons is ejected from the nucleus, is calculated using the Valencia model of Nieves et al.~\cite{nieves_2p2h_2013}. This is unchanged from the previous version of \textsc{neut}.
\section{Atmospheric Neutrino Oscillation Analysis}
\label{sec:oa_skonly}

Atmospheric neutrino data are analyzed via fit of binned MC counts to data. We bin atmospheric neutrino MC and data events using two-dimensional bins of reconstructed cosine zenith angle and momentum. The bin definitions for the \num{29} samples used in this analysis are listed in \tabref{purity}. The zenith angle bin definitions were updated for this analysis, and are discussed in \secref{oa_skonly:zbin}.

Systematic uncertainties are incorporated in the analysis as additional nuisance parameters which modify the nominal MC prediction in the fit. We estimate the effect of each systematic uncertainty by quantifying the change induced in each analysis bin due to $+1\sigma$ and $-1\sigma$ deviations from its nominal value. The systematic uncertainties considered in this analysis are described in \secref{oa_skonly:syst}.

\subsection{Zenith Angle Binning}
\label{sec:oa_skonly:zbin}

\begin{figure}
\includegraphics[width=1.0\linewidth]{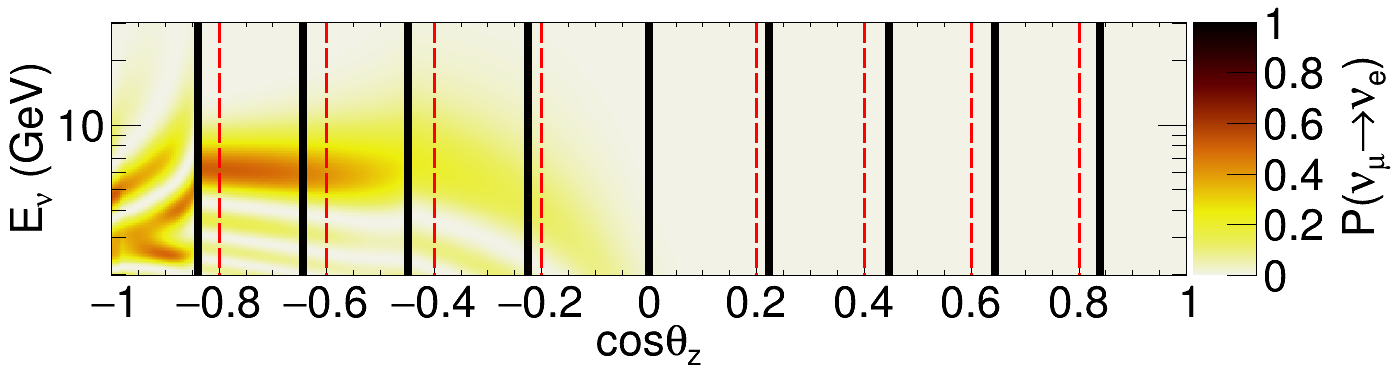}

\caption{Updated zenith angle bin edges for this analysis (black, solid lines) compared to equally-spaced bins (red, dashed lines). The updated bin edges are symmetric about zero and align with the matter-enhanced $\nu_{\mu}\to\nu_e$ resonance regions present assuming the PREM\@. The oscillation probabilities and corresponding color scale are the same as in \figref{oscillogram}, for neutrinos in the normal ordering.}
\label{fig:zenith_angle_bins}
\end{figure}

In the previous analysis, FC and PC events were binned into \num{10} evenly spaced cosine zenith angle bins on the interval [-1,1]. This separated the expected oscillation resonance region across two bins, reducing the signal-to-background ratio in samples sensitive to the neutrino mass ordering. In this analysis, the zenith angle bins for FC and PC events have been updated to more precisely cover the resonance region, demonstrated in \figref{zenith_angle_bins}. The updated bins are defined by the edges $-1$, $-0.839$, $-0.644$, $-0.448$, $-0.224$, $0$, $0.224$, $0.448$, $0.644$, $0.839$, and $1$. With the present statistics, adopting the updated bins results in a negligible effect on the expected sensitivity of this analysis to the mass ordering. However, the updated bins reduce the ambiguity as to whether or not events in the signal region occur in the expected zenith angle range or slightly outside of it.

\subsection{Systematic Uncertainties}
\label{sec:oa_skonly:syst}

The analysis includes \num{193} independent systematic uncertainty sources: \num{48} describe atmospheric neutrino flux and cross section effects common to all phases, and the remaining $29\times5$ describe reconstruction efficiencies which are separately estimated for each data taking phase.  Additional details on the formulation of the systematic uncertainties may be found in Ref.~\cite{twester_2023}.

\subsubsection{Flux and Cross Section Uncertainties}
\label{sec:analysis:syst:simulation}

The atmospheric neutrino flux uncertainties in this analysis are unchanged from the previous analysis~\cite{sk_atm_2018}. Two energy-dependent uncertainties scale the normalization of the flux above and below \qty{1}{\giga\ev}. Additionally, there are three energy-dependent uncertainties which modify the ratios of muon-to-electron, electron-to-anti-electron, and muon-to-anti-muon flavor neutrinos present in the atmospheric neutrino flux. An uncertainty based on the ratio of the Honda flux calculation with a modified kaon-to-pion ratio over the nominal model is also included~\cite{honda_2007}.

The effects of uncertainties on parameters in the CCQE and single-pion cross section models are computed by re-weighting MC events by the ratio of the double-differential cross section after $+ 1\sigma$ and $-1\sigma$ changes in each parameter. For CCQE events, $M_A^{\text{QE}}$ is taken to be $\qty{1.05(0.16)}{\giga\ev\per \c\squared}$. There are three parameters which control the single pion production cross section in the Rein-Sehgal model: the axial mass, $M_A^{\text{Res}}=\qty{0.95(0.15)}{\giga\ev\per \c\squared}$, the axial form factor coefficient, $C_{A}^{5} =\num{1.01(0.12)}$, and the isospin-$\frac{1}{2}$ background contribution scaling parameter, $I_{\frac{1}{2}}=\num{1.30(0.20)}$.

In addition to the $M_A^{\text{QE}}$ uncertainty, we place additional uncertainty on CCQE events due to differences between the LFG and RFG models. There are five uncertainties where the $1\sigma$ effect is computed by forming ratios between the predictions of the two models: the absolute event rate (both sub-GeV and multi-GeV), the shape of the $E_{\nu}$ dependence, and in the ratios of $\nu_{\mu}/\nu_e$ and $\bar{\nu}_{\mu}/\bar{\nu}_e$. These uncertainties are computed as a function of neutrino energy. The uncertainty on the 2p2h contribution to CCQE interactions is set to \qty{100}{\percent}, due to a lack of direct measurements, and is unchanged from the previous analysis.

An uncertainty on FSIs and SIs is implemented by re-computing the final states for MC events using multiple sets of the six \textsc{neut} FSI parameters. Each set consists of variations of one or more FSI parameters by their $\pm 1\sigma$ uncertainties as determined by a fit to external pion scattering from Ref.~\cite{t2k_2015}. The set which produces the largest change in classification outcome is taken as a conservative estimate of the FSI+SI uncertainty.

Two neutron-related systematic uncertainties are included in this analysis to account for the dependence on neutron production models in the neutron-based event selection used for \sk{IV} and \sk{V} data. Variations in these systematic uncertainties move events between samples with no tagged neutrons and samples with one or more tagged neutrons.  The largest of the two uncertainties originates from measurements of neutron production as a function of transverse momentum by T2K~\cite{akutsu_2020}. An additional neutron-related systematic uncertainty accounts for differences in neutron multiplicity predicted by the different neutrino interaction generators \textsc{neut} and \textsc{genie}~\cite{Andreopoulos:2009rq}.

In addition to model parameter uncertainties, we assign a \qty{20}{\percent} uncertainty on the ratio of NC to CC events and a \qty{25}{\percent} uncertainty on the tau neutrino CC cross section~\cite{sk_tau_2013}. The tau neutrino cross section uncertainty was estimated from a comparison between \textsc{neut} and the cross section model of Hagiwara et al.~\cite{hagiwara_2003}. 

\subsubsection{Reconstruction Uncertainties}
\label{sec:analysis:syst:det}

Ring reconstruction systematic uncertainties include efficiencies of ring-based particle identification (PID), ring counting, and identification of $\pi^0$ decays. The estimation of these systematic uncertainties is unchanged from the previous analysis, and is performed using a ``scale-and-shift'' procedure: First, MC events are labeled as either signal or background, where the particular label differs for each systematic uncertainty source. For example, the ring counting systematic uncertainty estimation considers events with a single charged particle with true momentum sufficient to produce a Cherenkov ring as signal, and events with multiple particles capable of producing rings as background. Similarly, the PID uncertainty estimation considers true $e$-like events as signal and true $\mu$-like events as background. Once labeled, the signal and background distributions of each reconstruction quantity, $x$, e.g., the ring-counting or PID likelihoods, are scaled and shifted by a linear function, $x' = \beta_0 + \beta_1 x$, and fit to a subset of atmospheric neutrino data. The signal and background distributions have independent $\beta$ parameters. The fitted $\beta$ parameters are then fluctuated according to their fitted uncertainties to generate toy data sets with random amounts of signal and background events. The maximum fractional change in signal purity observed in the toy data sets is taken as a conservative estimate of the uncertainty.

The scale-and-shift procedure is used to separately evaluate ring-related systematic uncertainties for events in the conventional and additional fiducial volume regions during each data-taking phase. The ring counting and PID systematic uncertainties are separately computed for each combination of $e$-like and $\mu$-like, and sub-GeV and multi-GeV events. As in the previous analysis, the scale-and-shift procedure is used to place an additional uncertainty on the normalization of $\nu_{\mu}$ contamination present in the $e$-like samples, estimated from the fitted fraction of $\nu_{\mu}$ events classified as $e$-like. We also use the scale-and-shift procedure to estimate the uncertainty associated with the multi-ring BDT: Two sources of systematic uncertainty are considered for multi-ring events based on the data-MC agreement in the input distributions used to train the BDT, and in the output BDT scores themselves. The changes in the event rate in each multi-ring sample obtained through fluctuating the input and output distributions according to their fitted uncertainty are taken as the $1\sigma$ effects.

As in the previous analysis, this analysis takes the maximum data-MC disagreement between the decay electron, $\pi^0$ mass, and stopping muon calibration sources during each SK phase, shown in \figref{escale}, as an estimate of the overall energy scale uncertainty. Variations in this uncertainty increase or decrease the reconstructed momenta of all MC events, causing these events to move between momentum bins. The energy scale uncertainty was newly estimated for the \sk{V} phase and is \qty{1.8}{\percent} in the conventional fiducial volume and \qty{2.0}{\percent} in the additional fiducial volume.

This analysis introduces a systematic uncertainty for the neutron tagging efficiency and false detection rate during the \sk{IV} and \sk{V} phases, which is factored into two components: detector conditions, and neutron travel distance. The contribution to the uncertainty due to detector conditions was estimated from comparisons between data from an AmBe source~\cite{first_ntag_2009} and MC events generated with variations in PMT gain, PMT dark rate, and water transparency within the ranges observed during the \sk{IV,V} phases. The neutron travel distance contribution to the uncertainty was estimated from an atmospheric neutrino data-MC comparison of the distance between the primary neutrino event vertex and reconstructed neutron capture vertex, convolved with the estimated capture efficiency as a function of the distance. The quadrature sum of these effects sets the overall neutron tagging uncertainty. Similarly to the neutron production uncertainties, variations in the tagging efficiency uncertainty move events between samples requiring no tagged neutrons and samples requiring one or more tagged neutron.

\begin{figure}
\includegraphics[width=1.0\linewidth]{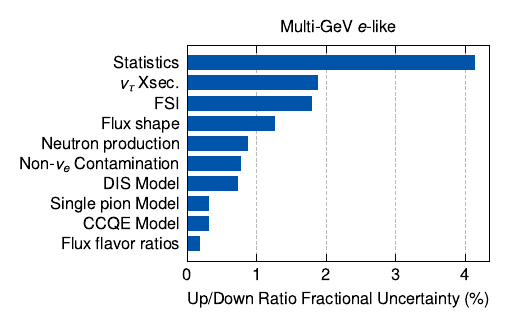}

\vspace{-1.3em}

\caption{Estimated fractional uncertainty on the ratio of upward-going, $\cos\theta_z<0$, versus downward-going, $\cos\theta_z>0$, events in the multi-GeV $e$-like samples (both single-ring and multi-ring) from statistical and systematic uncertainties. The fractional uncertainties are estimated by randomly fluctuating the nominal MC counts either by Poisson statistics for the ``Statistics'' entry, or by Gaussian fluctuations of one or more systematic uncertainties. Uncertainties on the up/down ratio in the multi-GeV $e$-like samples have the largest impact on the sensitivity to the neutrino mass ordering.}
\label{fig:up_down_largest_unc}
\end{figure}

The systematic uncertainties with the largest effect on the analysis' sensitivity to the neutrino mass ordering are those which can change the ratio of upward-going, $\cos\theta_z < 0$, versus downward-going, $\cos\theta_z >0$, events in the multi-GeV $e$-like samples. This is because the mass ordering signal region occurs in the upward-going bins of these samples, and downward-going neutrino events act as an un-oscillated data set which constrains uncertainties without zenith angle dependence independent of oscillation effects. \Figref{up_down_largest_unc} shows the estimated largest sources of uncertainty which affect the ratio of upward-versus-downward going events in the multi-GeV $e$-like samples. The largest source of uncertainty is statistics, indicating the mass ordering sensitivity is statistics-limited. The $\nu_{\tau}$ cross section uncertainty is the next-largest: Tau neutrino interactions at SK are almost exclusively upward-going since they are the result of $\nu_{\mu}\to\nu_{\tau}$ oscillations, so there are no downward-going tau neutrino events to constrain the $\nu_{\tau}$ cross section uncertainty. Additionally, tau neutrino interactions tend to be reconstructed as multi-GeV, multi-ring $e$-like events, which places them in the mass ordering signal region~\cite{li2018}. In contrast, uncertainties on the overall flux and cross section normalizations are less important for the mass ordering analysis, since these modify the predicted number of events independent of zenith angle.

\subsection{Fitting Method}
\label{sec:oa_skonly:fitting}

This work presents two fits of atmospheric neutrino data: First, we perform an ``atmospheric-only'' fit which measures $\dms{32}{,31}$, \sq{2}{3}, \dcp, \sq{1}{3}, and the neutrino mass ordering using the SK atmospheric neutrino data with no external constraints. Next, we analyze the same data including a constraint on \sq{1}{3}{} from reactor neutrino experiments.

In both fits, we evaluate data-MC agreement by computing a $\chi^2$ statistic at each point on a fixed grid of neutrino oscillation parameters, defined for each fit in \tabref{osc_grid}. Oscillation probabilities are applied to the MC events using the oscillation parameters at each point in the grid, and systematic uncertainty nuisance parameters are fit to determine the lowest $\chi^2$ value. The best-fit neutrino oscillation parameters are taken to be the grid point with the lowest $\chi^2$ value.

\begin{table}
\caption{The oscillation parameter grid definitions used in the three analyses presented in this work. The grids are evenly spaced, and include the minimum and maximum values. The grids are the same for both the normal and inverted mass orderings in all analyses. All three analyses treat \dms{2}{1} and \sq{1}{2} as constrained parameters.}
\begin{ruledtabular}
\begin{tabular}{lccc}
\textrm{Parameter}&
\textrm{Min.}&
\textrm{Max.}&
\textrm{Points}\\
\midrule
    \multicolumn{4}{l}{SK, Atmospheric Only} \\
    \hspace{1em}\sq{2}{3} & 0.3 & 0.775 & 20\\
    \hspace{1em}\dms{3}{2} (\qty{e-3}{\ev\squared}) & 1.2 & 3.6 & 25 \\ 
    \hspace{1em}\dcp{} (rad) & $-\pi$ & $\pi$ & 21 \\ 
    \hspace{1em}\sq{1}{3} & 0 & 0.075 & 16\\ \addlinespace
    \multicolumn{4}{l}{SK, \sq{1}{3} Constrained} \\
    \hspace{1em}\sq{2}{3} & 0.3 & 0.7 & 35\\
    \hspace{1em}\dms{3}{2} (\qty{e-3}{\ev\squared}) & 1.0 & 4.9 & 40 \\ 
    \hspace{1em}\dcp{} (rad) & $-\pi$ & $\pi$ & 37 \\ 
    \hspace{1em}\sq{1}{3}  & \multicolumn{2}{c}{\num{0.0220(0.0007)}} & 1 \\ 
\end{tabular}
\end{ruledtabular}
\label{tab:osc_grid}
\end{table}

The $\chi^2$ statistic is computed via a summation over $n$ bins assuming Poisson statistics, and with systematic uncertainty pull terms, $\epsilon_i$, which have units of $\sigma$. Each $\epsilon_i$ scales the effect of the $i^{\text{th}}$ systematic uncertainty, and is included in the $\chi^2$ calculation as an additional penalty term:
\begin{equation}
\label{eq:chi2}
\chi^2 = \sum_n \left(E_n - O_n + O_n \ln \frac{O_n}{E_n}\right) + \sum_i \epsilon_i^2 
\end{equation}

\noindent In \eqnref{chi2}, $n$ indexes each bin and $O_n$ are the observed counts in the $n^{\text{th}}$ bin. The expected counts $E_n$ are calculated from a nominal MC prediction $E_{n}^0$ scaled by the effect of systematic uncertainties. The effect of the $i^{\text{th}}$ systematic uncertainty is calculated as $\epsilon_i$ times the fractional change induced in the $n^{\text{th}}$ bin by a $1\sigma$ change, $f_{i,n}$:
\begin{equation}
    \label{eq:mc_expect}
    E_n = E_{n}^0 \left(1+ \sum_i f_{i,n} \epsilon_i\right)
\end{equation}

The constraint $\partial \chi^2 /\partial \epsilon_i = 0$ at the minimum $\chi^2$ value yields a system of equations which may be solved to find the configuration of $\epsilon_i$s that produce the smallest $\chi^2$ for a given MC prediction.

\begin{figure}
\includegraphics[width=1.0\linewidth]{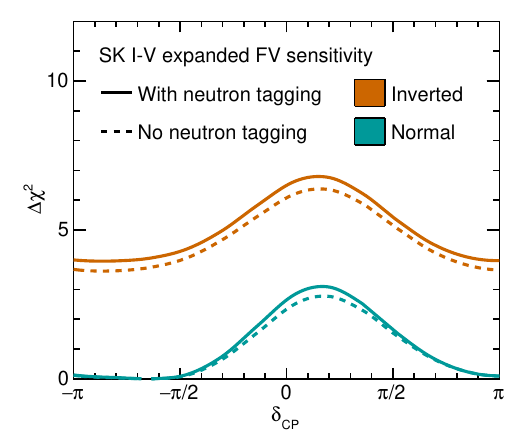}

\vspace{-1.3em}

\caption{1D $\Delta \chi^2$ sensitivity profile of \dcp{}  for the normal and inverted mass ordering scenarios, with (solid lines) and without (dashed lines) neutron-based event classification for \sk{IV} and \sk{V} data. The sensitivity is computed assuming the true neutrino oscillation parameters are the global best-fit parameters from~\cite{pdg2022} and the normal mass ordering. The parameters \dms{3}{2}, \sq{2}{3}, and \dcp{} are free parameters in the fit, while the other oscillation parameters are fixed.}
\label{fig:chi2_ntag}
\end{figure}

\begin{figure*}
\includegraphics[width=0.74\textwidth]{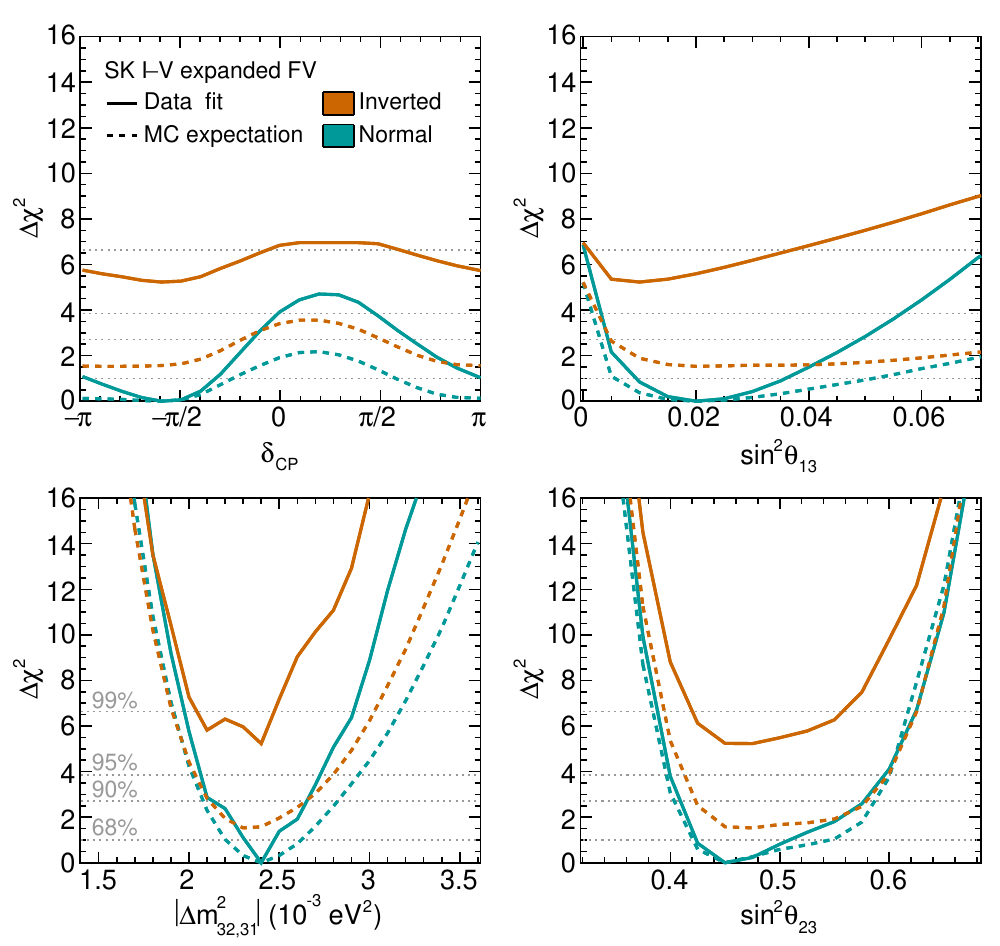}

\caption{1D $\Delta \chi^2$ profiles of oscillation parameters in the SK-only analysis with \sq{1}{3}{} treated as a free parameter. Solid lines correspond to the data fit result, while dashed lines correspond to the MC expectation at the data best-fit oscillation parameters, cf.\ \tabref{osc_grid}. Dashed lines show critical values of the $\chi^2$ distribution for 1 degree of freedom with lowest to highest corresponding to \qty{68}{\percent}, \qty{90}{\percent}, \qty{95}{\percent}, and \qty{99}{\percent} probabilities.}
\label{fig:skonly_q13_free}
\end{figure*}

\begin{figure*}
\includegraphics[width=0.8\textwidth]{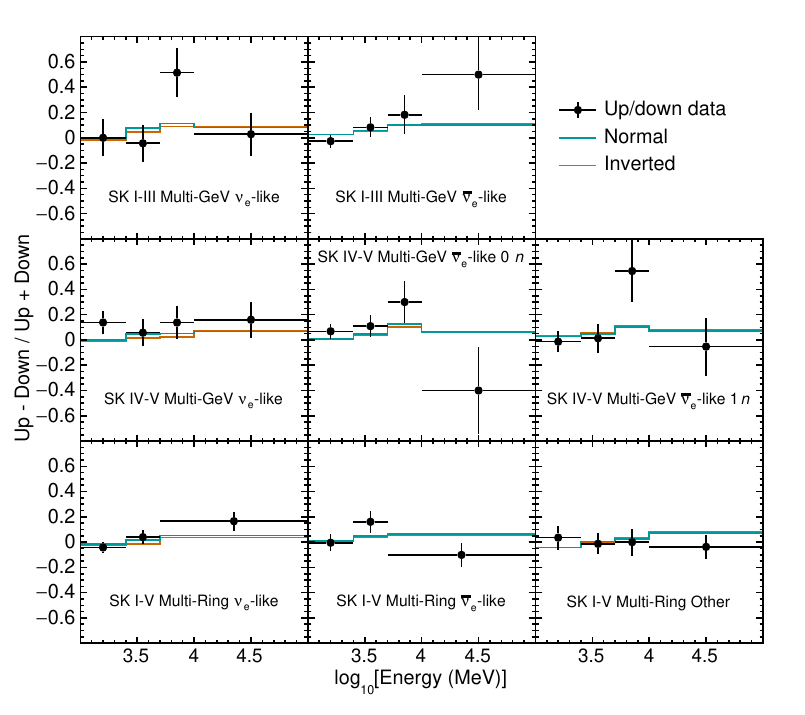}

\vspace{-1em}

\caption{Up-down asymmetry for multi-GeV $e$-like events. The $y$-axis is the asymmetry parameter, the ratio between the difference and sum of upward-going, $\cos\theta_{z} < 0.4$ and downward going, $\cos \theta_z > 0.4$ events. The $x$-axis is the reconstructed neutrino energy: For single ring events, the reconstructed energy is the visible energy of the ring assuming the reconstructed ring is an electron, while for multi-ring events, it is the total visible energy of the event. All error bars are statistical. MC lines for the normal and inverted orderings are drawn assuming the best-fit oscillation parameters of the SK + \sq{1}{3}-constrained analysis. \sk{IV-V} multi-GeV single-ring events are selected using the number of tagged neutrons, and so are separated from the \sk{I-III} multi-GeV single-ring samples.}
\label{fig:updown_asym}
\end{figure*}

\subsection{Mass Ordering Sensitivity with Neutron Tagging}
\label{sec:oa_skonly:ntag}

We studied the impact of using neutron captures on hydrogen with a tagging efficiency of \qty{26}{\percent} on the sensitivity to the mass ordering. As discussed in \secref{detector:selection:neutron}, tagged neutrons provide an additional handle during event reconstruction which improves the statistical separation of neutrinos and anti-neutrinos in SK versus only considering the number of decay electrons. This corresponds to an increased purity of correct-sign $\nu_{e}$ and $\bar{\nu}_{e}$ events in samples where sensitivity to the the mass  ordering is expected. The increased purity can be seen by comparing the FC multi-GeV single ring $e$-like samples listed in \tabref{purity}: The \sk{I,III} multi-GeV $\bar{\nu}_e$-like sample requiring no decay electron has a MC purity of \qty{34}{\percent} true $\bar{\nu}_e$ CC interactions, while the \sk{IV,V} multi-GeV $\bar{\nu}_e$-like sample requiring no decay electrons and one tagged neutron has a MC purity of \qty{46}{\percent} true $\bar{\nu}_e$ CC interactions.

\Figref{chi2_ntag} shows a comparison of the sensitivities obtained using the standard \num{19} samples discussed in \secref{detector:selection} for all SK phases versus the present configuration, which separates out the FC single-ring samples from the \sk{IV} and \sk{V} phases (approximately \qty{57}{\percent} of the total exposure) into additional samples which use neutron information. The figure demonstrates that using the neutron tagged samples improves the expected sensitivity to the mass ordering and \dcp{} by producing higher $\Delta \chi^2$ value [see \eqnref{chi2}] away from the true oscillation parameters.

\subsection{Results}
\label{sec:oa_skonly:results}

\subsubsection{Atmospheric-only Results}

Results in this section and the next are presented as $\Delta \chi^2$ contours of one or two oscillation parameters taken with respect to the global best-fit point, with and without constraints on \qq{1}{3}. The contours are profiled: We draw the minimum $\Delta \chi^2$ values among all other combinations of oscillation parameters scanned with one or two oscillation parameters fixed.

\Figref{skonly_q13_free} shows the one-dimensional (1D) $\Delta \chi^2$ contours in the atmospheric-only analyses for the fitted neutrino oscillation parameters \dcp, \sq{1}{3}, \dms{32}{,31}, and \sq{2}{3}, with respect to the best-fit point across both mass orderings. The normal ordering is preferred: The difference between the minimum $\chi^2$ in the inverted and normal orderings is $\dmo=\num[round-mode=places,round-precision=2]\skonlydmo$. The MC expectation for the mass ordering at the best-fit oscillation parameters is $\dmo=1.53$, indicated by dashed lines in \figref{skonly_q13_free}. The difference between the data result and the MC expectation is discussed in \secref{interp}.

Sensitivity to \dcp{} in the SK data originates from both sub-GeV and multi-GeV $e$-like samples, where values of \dcp{} near $-\pi/2$ indicate increased $\nu_{e}$ appearance and decreased $\bar{\nu}_e$ appearance relative to $\dcp=0$. The best-fit value of \dcp{} is \num[round-mode=places,round-precision=2]{\skonly[3,1]} in both orderings, signalling increased $\nu_e$ appearance. The constraints on \dcp{} and \sq{1}{3}{} assuming the inverted ordering are weaker, although consistent with the best-fit values assuming the normal ordering. These weaker constraints are an expected consequence of the freedom to adjust \sq{1}{3}{} and \dcp{} simultaneously, combined with the reduced anti-neutrino statistics relative to neutrino statistics in the atmospheric neutrino data.

The preferred values of \sq{1}{3}{} are \num[round-mode=places,round-precision=3]{\skonly[4,1]} and \num[round-mode=places,round-precision=3]{\skonlyi[4,1]} in the normal ordering and inverted ordering respectively. Both results are consistent with the reactor-preferred value of $\sq{1}{3}=0.0220$ at the \qty{90}{\percent} confidence level, and $\sq{1}{3}=0$ is disfavored in the normal ordering at the \qty{99}{\percent} level. The data favor the magnitude of the fitted squared mass difference $\lvert \dms{32}{,31}\lvert=\qty[round-mode=places,round-precision=2]{\skonly[1,1]e-3}{\ev\squared}$ in both orderings. The non-smooth behavior of the constraints on this parameter, especially evident in the inverted ordering fit, is a consequence of rapidly varying oscillation probabilities in the sub-GeV samples. Finally, the atmospheric neutrino data place the best-fit value of \sq{2}{3} in the lower octant, $\sq{2}{3}=\num[round-mode=places,round-precision=2]{\skonly[2,1]}$, although values in each octant are allowed at the \qty{68}{\percent} level.

As discussed in \secref{oscillation}, the combination of nonzero \sq{1}{3}{} and a normal neutrino mass ordering leads to electron neutrino appearance for upward-going multi-GeV events. We observe excess electron-flavor upward-going multi-GeV, single ring and multi-GeV, multi-ring events in the SK data. \Figref{updown_asym} shows a projection of the multi-GeV $e$-like samples as an up-down asymmetry:
\begin{equation}
    \text{Asymmetry} = \frac{\text{Up}-\text{Down}}{\text{Up}+\text{Down}},
\end{equation}
 
 \noindent where ``Up'' is the number of upward-going, $\cos \theta_z< -0.4$, events, and ``Down'' is the number of downward-going, $\cos \theta_z > 0.4$ events. The figure plots the asymmetry for these data as a function of reconstructed energy, and the expected asymmetry for the normal and inverted ordering scenarios assuming the best-fit oscillation parameters from the fit to all atmospheric neutrino data. The $\nu_{e}$-enhanced samples, multi-GeV $\nu_{e}$-like and multi-ring $\nu_{e}$-like, have the largest excesses relative to either ordering, and drive the preference for the normal mass ordering in the analysis.

 \begin{figure*}
\includegraphics[width=1.0\textwidth]{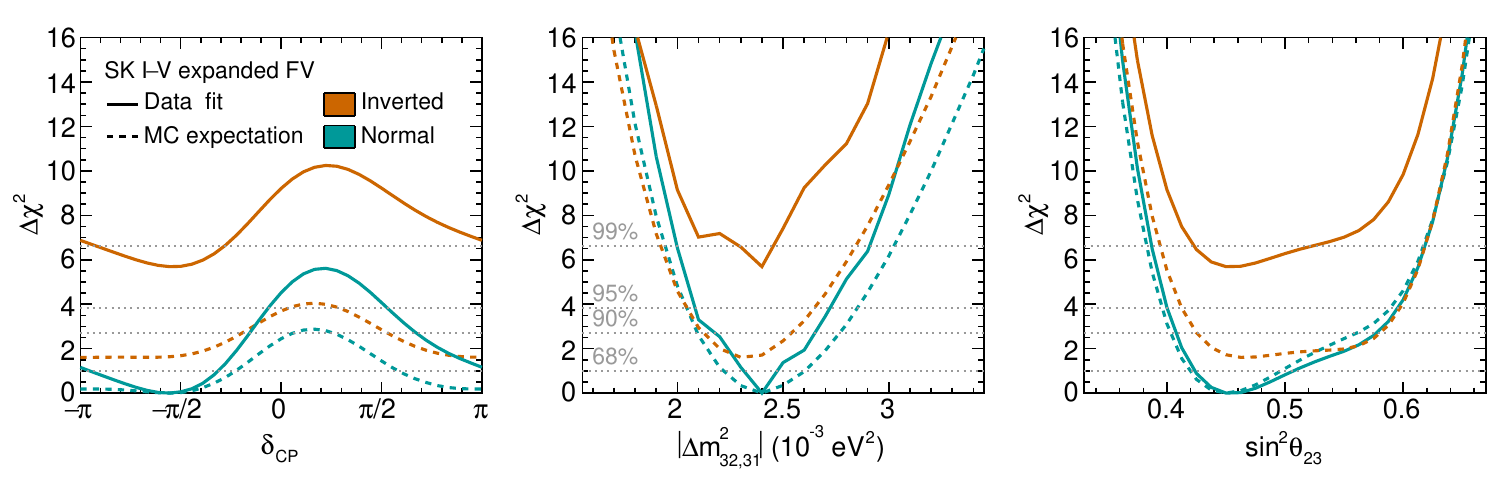}

\vspace{-1em}

\caption{1D $\Delta \chi^2$ profiles of oscillation parameters in the SK-only analysis with \sq{1}{3}{} constrained. Solid lines correspond to the data fit result, while dashed lines correspond to the MC expectation at the data best-fit oscillation parameters, cf.\ \tabref{osc_grid}. Dashed lines show critical values of the $\chi^2$ distribution for 1 degree of freedom corresponding to \qty{68}{\percent}, \qty{90}{\percent}, \qty{95}{\percent}, and \qty{99}{\percent} probabilities.}
\label{fig:skonly_1d}
\end{figure*}

\subsubsection{Results with Reactor Constraints on \texorpdfstring{\sq{1}{3}}{sin2Q13}}
\label{sec:oa_skonly:results:analysis_q13}

\Figref{skonly_1d} shows the 1D $\Delta \chi^2$ profiles for the fitted neutrino oscillation parameters assuming the constraint $\sq{1}{3}=0.0220\pm0.0007$ from reactor antineutrino disappearance experiments~\cite{pdg2022}. The constraint on \sq{1}{3}{} is incorporated by introducing an additional systematic uncertainty for this fit, where the $1\sigma$ effect is defined as the change induced by varying $\sq{1}{3}$ by its measured $1\sigma$ uncertainty.

The best-fit value of $\dcp$ in both the normal and inverted orderings for the fit with \sq{1}{3}{} constrained is \num[round-mode=places,round-precision=2]{\skqfix[3,1]}, which is consistent with the atmospheric-only analysis at the $1\sigma$ level. This fit also finds improved constraints on \dcp{} in the inverted ordering for values near $\pi/2$: The constraint on \sq{1}{3}{} fixes the effect size of the mass ordering, such that the separate modifications to $\nu_e$ appearance from \dcp{} are more readily resolved.

In this fit, the preference for the normal ordering increases to $\dmo=\num[round-mode=places,round-precision=2]{\skqfixdmo}$. This improvement is consistent with the observed preference for smaller value of \sq{1}{3} in the inverted ordering fit with \sq{1}{3} free: The $\chi^2$ value in the inverted ordering increases with the added constraint, while the $\chi^2$ value in the normal ordering remains similar to the result without the constraint.

\begin{figure}
\includegraphics[width=1.0\linewidth]{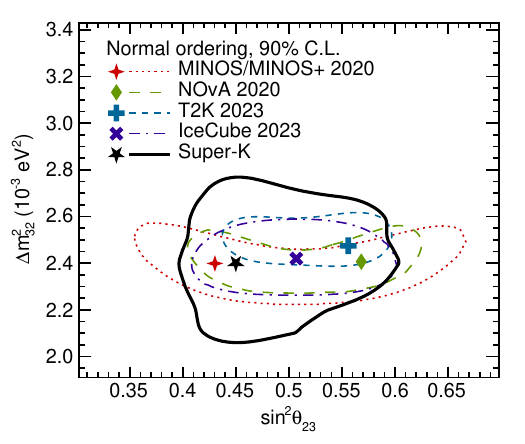}

\vspace{-1em}

\caption{2D constant $\Delta \chi^2$ contours of neutrino oscillation parameters $\Delta  m^2_{32}$ and  \sq{2}{3}{} for the normal mass ordering. Contours are drawn for a \qty{90}{\percent} critical $\chi^2$ value assuming 2 degrees of freedom, with the $\Delta \chi^2$ computed for each experiment with respect to the best-fit point in the normal mass  ordering. The Super-K contour shows the result of this analysis, and other contours are adapted from publications by MINOS+~\cite{minos_2020}, NOvA~\cite{alex_himmel_2020_4142045}, T2K~\cite{t2k_2023}, and IceCube~\cite{icecube_2023}. Best-fit points are indicated with markers for each experiment.}
\label{fig:experiments}
\end{figure}

\Figref{experiments} shows the comparison of the results of the \qq{1}{3}-constrained analysis of SK atmospheric neutrino data with other contemporary results from MINOS/MINOS+~\cite{minos_2020}, NOvA~\cite{nova_2022}, T2K~\cite{t2k_2023}, and IceCube~\cite{icecube_2023}. As can be seen from the contours, SK atmospheric neutrino data are consistent with the other experiments at the \qty{90}{\percent} level. While the atmospheric neutrino data find a best-fit value of \sq{2}{3}{} in the lower octant, we note that the previous publication found a best-fit value in the upper octant \cite{sk_atm_2018}, and that value of \sq{2}{3}{} in both octants are allowed at the $1\sigma$ level. 

\begin{table*}
\sisetup{table-alignment-mode=format, table-number-alignment=center, round-mode=places, round-precision=2}
\setlength{\tabcolsep}{6pt}
 \caption{Best-fit neutrino oscillation parameters from the analyses presented in this work. The uncertainties on each oscillation parameter are the $\pm 1\sigma$ allowed regions assuming a $\chi^2$ distribution with one degree of freedom. The second-to-last column shows the total $\chi^2$. Both analyses have \num[round-precision=0]{930} bins.}


\begin{ruledtabular}
\begin{tabular}{@{}
lc
S[table-format = 1.2,table-space-text-post={$\phantom{}^{+1.11}$}]
S[table-format = 1.2,table-space-text-post={$\phantom{}^{+1.11}$}]
S[table-format = 1.3,round-precision=3,table-space-text-post={$\phantom{}^{+1.111}$}]
S[table-format = +1.2,table-space-text-post={$\phantom{}^{+1.11}$}]
S[table-format = 4.2,round-mode=places,round-precision=2]
c
@{}}
{Fit result}
& {Ordering}
& {$\lvert\dms{3}{2,31}\lvert$}
& {$\sin^2 \theta_{23}$}
& {$\sin^2 \theta_{13}$}
& {\dcp}
& {$\chi^2$}
& {\dmo} \\
& & {\footnotesize{(\qty[round-mode=none]{e-3}{\ev\squared})}} & & & {\footnotesize $(-\pi,\pi)$} & & \\
\midrule
{\multirow{2}*[-0.5ex]{SK, Atmospheric Only}} & 
{Normal} &
\skonly[1,1]$\phantom{}^{+\skonly[1,2]}_{-\skonly[1,3]}$ &
\skonly[2,1]$\phantom{}^{+\num{\skonly[2,2]}}_{-\num{\skonly[2,3]}}$ &
\skonly[4,1]$\phantom{}^{+\num{\skonly[4,2]}}_{-\num{\skonly[4,3]}}$ &
\skonly[3,1]$\phantom{}^{+\num{\skonly[3,2]}}_{-\num{\skonly[3,3]}}$ &
\skonly[5] & 
\multirow{2}*[-0.5ex]{\tablenum[table-format = 2.2]{\skonlydmo}}\\ \addlinespace
& {Inverted} &
\skonlyi[1,1]$\phantom{}^{+\skonlyi[1,2]}_{-\skonlyi[1,3]}$ &
\skonlyi[2,1]$\phantom{}^{+\num{\skonlyi[2,2]}}_{-\num{\skonlyi[2,3]}}$ &
\skonlyi[4,1]$\phantom{}^{+\skonlyi[4,2]}_{-\skonlyi[4,3]}$ &
\skonlyi[3,1]$\phantom{}^{+\num{\skonlyi[3,2]}}_{-\num{\skonlyi[3,3]}}$ &
\skonlyi[5]\\ \addlinespace\addlinespace
{\multirow{2}*[-0.5ex]{SK, \sq{1}{3}{} Constrained}} & 
{Normal} &
\skqfix[1,1]$\phantom{}^{+\skqfix[1,2]}_{-\skqfix[1,3]}$ &
\skqfix[2,1]$\phantom{}^{+\num{\skqfix[2,2]}}_{-\num{\skqfix[2,3]}}$ &
{--} &
\skqfix[3,1]$\phantom{}^{+\num{\skqfix[3,2]}}_{-\num{\skqfix[3,3]}}$ &
\skqfix[5] & 
\multirow{2}*[-0.5ex]{\tablenum[table-format = 2.2]{\skqfixdmo}}\\ \addlinespace
& {Inverted} &
\skqfixi[1,1]$\phantom{}^{+\skqfixi[1,2]}_{-\skqfixi[1,3]}$ &
\skqfixi[2,1]$\phantom{}^{+\num{\skqfixi[2,2]}}_{-\num{\skqfixi[2,3]}}$ &
{--} &
\skqfixi[3,1]$\phantom{}^{+\num{\skqfixi[3,2]}}_{-\num{\skqfixi[3,3]}}$ &
\skqfixi[5]\\
\end{tabular}
\end{ruledtabular}
 \label{tab:fit_results}
\end{table*}

\section{Interpretation}
\label{sec:interp}

\Tabref{fit_results} summarizes the fit results of the analyses presented in this work. In both analyses, the normal ordering is preferred, and the best-fit oscillation parameters predict weaker sensitivities to the neutrino mass ordering than the observed \dmo. To quantify the significance of the mass ordering preference from the fit results, we generated ensembles of toy data sets to produce distribution of the \dmo{} statistic\footnote{Because the two mass ordering scenarios are not nested hypotheses, taking the square root of \dmo to estimate the significance by invoking Wilks' theorem is not recommended~\cite{algeri2020}.}. Each toy data set consists of fluctuated counts according to each bin's statistical uncertainty which are scaled by randomly sampling the systematic uncertainty coefficients. Ensembles were generated assuming both the normal and inverted mass orderings and with oscillation parameters fixed at the best-fit points. We fit each toy data set in each ordering with \dms{3}{2}, \sq{2}{3}, and \dcp{} as free parameters to compute \dmo.

\begin{figure}
\includegraphics[width=1.0\linewidth]{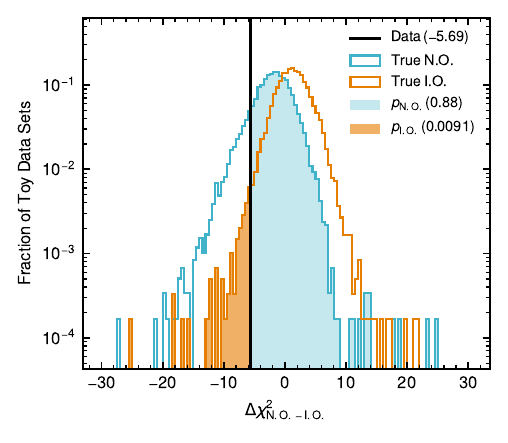}

\vspace{-2em}

\caption{\label{fig:pval}Distribution of the mass ordering preference statistic, \dmo, for ensembles of simulated data sets, assuming either the normal or inverted mass orderings. The data result from the atmospheric analysis with \sq{1}{3}-constrained analysis is shown as the vertical black line. The blue and orange histograms indicate the distribution of this statistic for toy data sets assuming the normal and inverted ordering respectively. The filled areas to the left of the data result for inverted toy data sets and to the right of the data result for normal toy data sets indicate the $p$-values.}
\end{figure}

\Figref{pval} shows the distribution of \dmo{} compared with the data fit result for the atmospheric analysis with \sq{1}{3}{} constrained. The probability of observing a more extreme result than the data (the $p$-value) is given by the area to the right of the data line in the normal ordering scenario, and by the area to the left of the data line in the inverted ordering scenario. While the figure shows the $p$-value determined from simulated data sets for the inverted mass ordering is \num[round-mode=places,round-precision=4]{\skqfixp[1,2]}, we note that with the present SK statistics, the expected sensitivity remains weak for rejecting either ordering. Indeed, the $p$-value for the data result within the normal ordering, \num[round-mode=places,round-precision=2]{\skqfixp[1,1]}, is not especially likely either. For the situation in which the data must select between two mutually-exclusive hypotheses, the \cls{} method~\cite{cls_2002} provides an estimate of the $p$-value that considers simultaneous agreement from both hypotheses:
\begin{equation}
\label{eq:cls}
\cls = \frac{p_{\text{I.O.}}}{1-p_{\text{N.O.}}},
\end{equation}

\noindent where $p_{\text{N.O.}}$ and $p_{\text{I.O.}}$ refer to the $p$-values in the normal or inverted ordering. This prescription decreases the significance for rejecting the inverted hypothesis proportional to the simultaneous significance of accepting the normal ordering hypothesis. The \cls{} for the atmospheric analysis with \sq{1}{3} constrained is \num[round-mode=places,round-precision=3]{\skqfixcls[1]}, corresponding to a rejection of the inverted mass ordering at the \qty[round-mode=places,round-precision=1]{\skqfixpref}{\percent} confidence level. This number is similar to the previous SK result, $\cls=0.070$~\cite{sk_atm_2018}, despite originating from a larger $\Delta \chi^2$, \num[round-mode=places,round-precision=2]{\skqfixdmo} versus $4.33$. While the mass ordering sensitivity and data result both increased for this analysis, the probability of obtaining the data result simultaneously decreased in both orderings, resulting in a similar \cls{} value.

The $p$-value obtained from toy data sets depends on the choice of oscillation parameters. While the atmospheric analyses places \sq{2}{3}{} in the lower octant, values of \sq{2}{3}{} spanning both octants are allowed at the $1\sigma$ level. Larger values of \sq{2}{3}{} and \dcp{} values near $-\pi/2$ increase the sensitivity of SK for rejecting the incorrect mass ordering since they enhance the $\nu_{\mu}\to\nu_e$ signal. Accordingly, the mass ordering is more difficult to resolve for smaller values of \sq{2}{3}{} and values of \dcp{} near $\pi/2$. To demonstrate the dependence of \cls{} outcomes on the choice of oscillation parameters, we repeated the generation of toy data sets for configurations of oscillation parameters which maximize and minimize the expected sensitivity to rejecting the incorrect mass ordering and are allowed at the \qty{90}{\percent} confidence level. The range of \cls{} values obtained in the atmospheric fit with \sq{1}{3} constrained span \qtyrange[round-mode=places,round-precision=3]{\skqfixcls[2]}{\skqfixcls[3]}.. We observe that upper-octant values of \sq{2}{3}{} predict larger\dmo{} values which are closer to the observed data result. We anticipate that better constraints on the \sq{2}{3}{} octant will reduce the difference between the \dmo{} values expected from MC and obtained from data.

\section{Conclusion}
\label{sec:interp:conclusion}
{
\sisetup{round-mode=places,round-precision=2}We analyzed \num[round-mode=places,round-precision=1]{\sklt} live-days of atmospheric neutrino data collected with the Super-Kamiokande experiment operating with pure water and an expanded fiducial volume. An event selection using tagged neutron information was used to enhance the statistical separation of neutrino and anti-neutrino data, thereby increasing the sensitivity to the neutrino mass ordering. An analysis of SK  data with constraints on \sq{1}{3}{} measures the oscillation parameters to be $\dms{3}{2}=\skqfix[1,1]^{+\skqfix[1,2]}_{-\skqfix[1,3]}\,\unit{\ev\squared}$, $\sq{2}{3}=\num{\skqfix[2,1]}^{+\num{\skqfix[2,2]}}_{-\num{\skqfix[2,3]}}$, and $\dcp=\num{\skqfix[3,1]}^{+\num{\skqfix[3,2]}}_{-\num{\skqfix[3,3]}}$. The analysis prefers the normal ordering over the inverted ordering at the \qty[round-mode=places,round-precision=1]{\skqfixpref}{\percent} confidence level.  We anticipate improvements to the mass ordering sensitivity in future analyses of SK atmospheric data which include gadolinium-enhanced neutron tagging for enhanced neutrino and anti-neutrino separation.

This work is accompanied by a data release~\cite{sk_data_release_2023}. The data release contains the 1D and 2D contours of fitted oscillation parameters at the \qty[round-mode=places, round-precision=0]{68}{\percent} and \qty[round-mode=places, round-precision=0]{90}{\percent} confidence levels from the analyses described in \secref{oa_skonly}, and listings of the data and MC counts in the \num[round-mode=places, round-precision=0]{930} atmospheric neutrino bins used throughout this work. Summary statistics of the MC neutrino energies, directions, and flavors are provided for each bin.

\begin{acknowledgments}
We gratefully acknowledge cooperation of the Kamioka Mining and Smelting Company.
The Super-Kamiokande experiment was built and has been operated with funding from the
Japanese Ministry of Education, Science, Sports and Culture, 
and the U.S. Department of Energy.

We gratefully acknowledge the cooperation of the Kamioka Mining and Smelting Company.
The Super-Kamiokande experiment has been built and operated from funding by the 
Japanese Ministry of Education, Culture, Sports, Science and Technology, the U.S.
Department of Energy, and the U.S. National Science Foundation. Some of us have been 
supported by funds from the National Research Foundation of Korea (NRF-2009-0083526 
and NRF 2022R1A5A1030700) funded by the Ministry of Science, 
Information and Communication Technology (ICT), the Institute for 
Basic Science (IBS-R016-Y2), and the Ministry of Education (2018R1D1A1B07049158,
2021R1I1A1A01042256, the Japan Society for the Promotion of Science, the National
Natural Science Foundation of China under Grants No.11620101004, the Spanish Ministry of Science, 
Universities and Innovation (grant PGC2018-099388-B-I00), the Natural Sciences and 
Engineering Research Council (NSERC) of Canada, the Scinet and Westgrid consortia of
Compute Canada, the National Science Centre (UMO-2018/30/E/ST2/00441 and UMO-2022/46/E/ST2/00336) and the Ministry of Education and Science (2023/WK/04), Poland,
the Science and Technology Facilities Council (STFC) and
Grid for Particle Physics (GridPP), UK, the European Union's 
Horizon 2020 Research and Innovation Programme under the Marie Sklodowska-Curie grant
agreement no.754496, H2020-MSCA-RISE-2018 JENNIFER2 grant agreement no.822070, 
H2020-MSCA-RISE-2019 SK2HK grant agreement no.872549, and H2020-MSCA-GF-2019 HYPER-KOD grant agreement no.892264.

\end{acknowledgments}

\bibliographystyle{apsrev4-2}
\bibliography{sk_atm_2023}

\providecommand{\noopsort}[1]{}\providecommand{\singleletter}[1]{#1}%
\begin{thebibliography}{59}%
\makeatletter
\providecommand \@ifxundefined [1]{%
 \@ifx{#1\undefined}
}%
\providecommand \@ifnum [1]{%
 \ifnum #1\expandafter \@firstoftwo
 \else \expandafter \@secondoftwo
 \fi
}%
\providecommand \@ifx [1]{%
 \ifx #1\expandafter \@firstoftwo
 \else \expandafter \@secondoftwo
 \fi
}%
\providecommand \natexlab [1]{#1}%
\providecommand \enquote  [1]{``#1''}%
\providecommand \bibnamefont  [1]{#1}%
\providecommand \bibfnamefont [1]{#1}%
\providecommand \citenamefont [1]{#1}%
\providecommand \href@noop [0]{\@secondoftwo}%
\providecommand \href [0]{\begingroup \@sanitize@url \@href}%
\providecommand \@href[1]{\@@startlink{#1}\@@href}%
\providecommand \@@href[1]{\endgroup#1\@@endlink}%
\providecommand \@sanitize@url [0]{\catcode `\\12\catcode `\$12\catcode
  `\&12\catcode `\#12\catcode `\^12\catcode `\_12\catcode `\%12\relax}%
\providecommand \@@startlink[1]{}%
\providecommand \@@endlink[0]{}%
\providecommand \url  [0]{\begingroup\@sanitize@url \@url }%
\providecommand \@url [1]{\endgroup\@href {#1}{\urlprefix }}%
\providecommand \urlprefix  [0]{URL }%
\providecommand \Eprint [0]{\href }%
\providecommand \doibase [0]{https://doi.org/}%
\providecommand \selectlanguage [0]{\@gobble}%
\providecommand \bibinfo  [0]{\@secondoftwo}%
\providecommand \bibfield  [0]{\@secondoftwo}%
\providecommand \translation [1]{[#1]}%
\providecommand \BibitemOpen [0]{}%
\providecommand \bibitemStop [0]{}%
\providecommand \bibitemNoStop [0]{.\EOS\space}%
\providecommand \EOS [0]{\spacefactor3000\relax}%
\providecommand \BibitemShut  [1]{\csname bibitem#1\endcsname}%
\let\auto@bib@innerbib\@empty
\bibitem [{\citenamefont {Pontecorvo}(1962)}]{ponte1952}%
  \BibitemOpen
  \bibfield  {author} {\bibinfo {author} {\bibfnamefont {B.}~\bibnamefont
  {Pontecorvo}},\ }\href@noop {} {\bibfield  {journal} {\bibinfo  {journal}
  {Sov. Phys. JETP}\ }\textbf {\bibinfo {volume} {26}},\ \bibinfo {pages} {984}
  (\bibinfo {year} {1962})}\BibitemShut {NoStop}%
\bibitem [{\citenamefont {Maki}\ \emph {et~al.}(1962)\citenamefont {Maki},
  \citenamefont {Nakagawa},\ and\ \citenamefont {Sakata}}]{mns1962}%
  \BibitemOpen
  \bibfield  {author} {\bibinfo {author} {\bibfnamefont {Z.}~\bibnamefont
  {Maki}}, \bibinfo {author} {\bibfnamefont {M.}~\bibnamefont {Nakagawa}},\
  and\ \bibinfo {author} {\bibfnamefont {S.}~\bibnamefont {Sakata}},\ }\href
  {https://doi.org/10.1143/PTP.28.870} {\bibfield  {journal} {\bibinfo
  {journal} {Prog. Theor. Phys.}\ }\textbf {\bibinfo {volume} {28}},\ \bibinfo
  {pages} {870} (\bibinfo {year} {1962})}\BibitemShut {NoStop}%
\bibitem [{\citenamefont {Abe}\ \emph {et~al.}(2021)\citenamefont {Abe} \emph
  {et~al.}}]{t2k_2021}%
  \BibitemOpen
  \bibfield  {author} {\bibinfo {author} {\bibfnamefont {K.}~\bibnamefont
  {Abe}} \emph {et~al.} (\bibinfo {collaboration} {T2K Collaboration}),\
  }\bibfield  {journal} {\bibinfo  {journal} {Phys. Rev. D}\ }\textbf {\bibinfo
  {volume} {103}},\ \href {https://doi.org/10.1103/physrevd.103.112008}
  {10.1103/physrevd.103.112008} (\bibinfo {year} {2021})\BibitemShut {NoStop}%
\bibitem [{\citenamefont {Acero}\ \emph {et~al.}(2022)\citenamefont {Acero}
  \emph {et~al.}}]{nova_2022}%
  \BibitemOpen
  \bibfield  {author} {\bibinfo {author} {\bibfnamefont {M.}~\bibnamefont
  {Acero}} \emph {et~al.} (\bibinfo {collaboration} {NOvA Collaboration}),\
  }\bibfield  {journal} {\bibinfo  {journal} {Phys. Rev. D}\ }\textbf {\bibinfo
  {volume} {106}},\ \href {https://doi.org/10.1103/physrevd.106.032004}
  {10.1103/physrevd.106.032004} (\bibinfo {year} {2022})\BibitemShut {NoStop}%
\bibitem [{\citenamefont {Abe}\ \emph {et~al.}(2018)\citenamefont {Abe} \emph
  {et~al.}}]{sk_atm_2018}%
  \BibitemOpen
  \bibfield  {author} {\bibinfo {author} {\bibfnamefont {K.}~\bibnamefont
  {Abe}} \emph {et~al.} (\bibinfo {collaboration} {Super-Kamiokande
  Collaboration}),\ }\href {https://doi.org/10.1103/PhysRevD.97.072001}
  {\bibfield  {journal} {\bibinfo  {journal} {Phys. Rev. D}\ }\textbf {\bibinfo
  {volume} {97}},\ \bibinfo {pages} {072001} (\bibinfo {year}
  {2018})}\BibitemShut {NoStop}%
\bibitem [{\citenamefont {Adey}\ \emph {et~al.}(2018)\citenamefont {Adey} \emph
  {et~al.}}]{dayabay_2018}%
  \BibitemOpen
  \bibfield  {author} {\bibinfo {author} {\bibfnamefont {D.}~\bibnamefont
  {Adey}} \emph {et~al.} (\bibinfo {collaboration} {Daya Bay Collaboration}),\
  }\bibfield  {journal} {\bibinfo  {journal} {Phys. Rev. Lett.}\ }\textbf
  {\bibinfo {volume} {121}},\ \href
  {https://doi.org/10.1103/physrevlett.121.241805}
  {10.1103/physrevlett.121.241805} (\bibinfo {year} {2018})\BibitemShut
  {NoStop}%
\bibitem [{\citenamefont {Shin}\ \emph {et~al.}(2020)\citenamefont {Shin} \emph
  {et~al.}}]{reno_2020}%
  \BibitemOpen
  \bibfield  {author} {\bibinfo {author} {\bibfnamefont {C.~D.}\ \bibnamefont
  {Shin}} \emph {et~al.} (\bibinfo {collaboration} {RENO Collaboration}),\
  }\href {https://doi.org/10.1007/JHEP04(2020)029} {\bibfield  {journal}
  {\bibinfo  {journal} {JHEP}\ }\textbf {\bibinfo {volume} {2020}}\bibinfo
  {number} { (4)},\ \bibinfo {pages} {29}}\BibitemShut {NoStop}%
\bibitem [{\citenamefont {de~Kerret}\ \emph {et~al.}(2020)\citenamefont
  {de~Kerret} \emph {et~al.}}]{doublechooz_2020}%
  \BibitemOpen
\bibfield  {number} {  }\bibfield  {author} {\bibinfo {author} {\bibfnamefont
  {H.}~\bibnamefont {de~Kerret}} \emph {et~al.} (\bibinfo {collaboration}
  {Double-Chooz Collaboration}),\ }\href
  {https://doi.org/10.1038/s41567-020-0831-y} {\bibfield  {journal} {\bibinfo
  {journal} {Nature Physics}\ }\textbf {\bibinfo {volume} {16}},\ \bibinfo
  {pages} {558–564} (\bibinfo {year} {2020})}\BibitemShut {NoStop}%
\bibitem [{\citenamefont {Abe}\ \emph {et~al.}(2020)\citenamefont {Abe} \emph
  {et~al.}}]{t2knature_2020}%
  \BibitemOpen
  \bibfield  {author} {\bibinfo {author} {\bibfnamefont {K.}~\bibnamefont
  {Abe}} \emph {et~al.} (\bibinfo {collaboration} {T2K Collaboration}),\ }\href
  {https://doi.org/10.1038/s41586-020-2177-0} {\bibfield  {journal} {\bibinfo
  {journal} {Nature}\ }\textbf {\bibinfo {volume} {580}},\ \bibinfo {pages}
  {339} (\bibinfo {year} {2020})}\BibitemShut {NoStop}%
\bibitem [{\citenamefont {Ahmad}\ \emph {et~al.}(2001)\citenamefont {Ahmad},
  \citenamefont {Allen} \emph {et~al.}}]{sno_2001}%
  \BibitemOpen
  \bibfield  {author} {\bibinfo {author} {\bibfnamefont {Q.~R.}\ \bibnamefont
  {Ahmad}}, \bibinfo {author} {\bibfnamefont {R.~C.}\ \bibnamefont {Allen}},
  \emph {et~al.} (\bibinfo {collaboration} {SNO Collaboration}),\ }\href
  {https://doi.org/10.1103/PhysRevLett.87.071301} {\bibfield  {journal}
  {\bibinfo  {journal} {Phys. Rev. Lett.}\ }\textbf {\bibinfo {volume} {87}},\
  \bibinfo {pages} {071301} (\bibinfo {year} {2001})}\BibitemShut {NoStop}%
\bibitem [{\citenamefont {Fukuda}\ \emph {et~al.}(2001)\citenamefont {Fukuda}
  \emph {et~al.}}]{sk_2001}%
  \BibitemOpen
  \bibfield  {author} {\bibinfo {author} {\bibfnamefont {S.}~\bibnamefont
  {Fukuda}} \emph {et~al.} (\bibinfo {collaboration} {Super-Kamiokande
  Collaboration}),\ }\href {https://doi.org/10.1103/PhysRevLett.86.5651}
  {\bibfield  {journal} {\bibinfo  {journal} {Phys. Rev. Lett.}\ }\textbf
  {\bibinfo {volume} {86}},\ \bibinfo {pages} {5651} (\bibinfo {year}
  {2001})}\BibitemShut {NoStop}%
\bibitem [{\citenamefont {Wolfenstein}(1978)}]{wolfenstein_1978}%
  \BibitemOpen
  \bibfield  {author} {\bibinfo {author} {\bibfnamefont {L.}~\bibnamefont
  {Wolfenstein}},\ }\href {https://doi.org/10.1103/PhysRevD.17.2369} {\bibfield
   {journal} {\bibinfo  {journal} {Phys. Rev. D}\ }\textbf {\bibinfo {volume}
  {17}},\ \bibinfo {pages} {2369} (\bibinfo {year} {1978})}\BibitemShut
  {NoStop}%
\bibitem [{\citenamefont {Mikheev}\ and\ \citenamefont
  {Smirnov}(1985)}]{ms_1985}%
  \BibitemOpen
  \bibfield  {author} {\bibinfo {author} {\bibfnamefont {S.~P.}\ \bibnamefont
  {Mikheev}}\ and\ \bibinfo {author} {\bibfnamefont {A.~Y.}\ \bibnamefont
  {Smirnov}},\ }\bibfield  {journal} {\bibinfo  {journal} {Sov. J. Nucl. Phys.
  (Engl. Transl.)}\ }\textbf {\bibinfo {volume} {42:6}},\ \href
  {https://www.osti.gov/biblio/5714592} {} (\bibinfo {year} {1985})\BibitemShut
  {NoStop}%
\bibitem [{\citenamefont {Dziewonski}\ and\ \citenamefont
  {Anderson}(1981)}]{prem1981}%
  \BibitemOpen
  \bibfield  {author} {\bibinfo {author} {\bibfnamefont {A.~M.}\ \bibnamefont
  {Dziewonski}}\ and\ \bibinfo {author} {\bibfnamefont {D.~L.}\ \bibnamefont
  {Anderson}},\ }\href
  {https://doi.org/https://doi.org/10.1016/0031-9201(81)90046-7} {\bibfield
  {journal} {\bibinfo  {journal} {Phys. Earth Planet. Inter.}\ }\textbf
  {\bibinfo {volume} {25}},\ \bibinfo {pages} {297} (\bibinfo {year}
  {1981})}\BibitemShut {NoStop}%
\bibitem [{\citenamefont {Barger}\ \emph {et~al.}(1980)\citenamefont {Barger},
  \citenamefont {Whisnant}, \citenamefont {Pakvasa},\ and\ \citenamefont
  {Phillips}}]{barger}%
  \BibitemOpen
  \bibfield  {author} {\bibinfo {author} {\bibfnamefont {V.}~\bibnamefont
  {Barger}}, \bibinfo {author} {\bibfnamefont {K.}~\bibnamefont {Whisnant}},
  \bibinfo {author} {\bibfnamefont {S.}~\bibnamefont {Pakvasa}},\ and\ \bibinfo
  {author} {\bibfnamefont {R.~J.~N.}\ \bibnamefont {Phillips}},\ }\href
  {https://doi.org/10.1103/PhysRevD.22.2718} {\bibfield  {journal} {\bibinfo
  {journal} {Phys. Rev. D}\ }\textbf {\bibinfo {volume} {22}},\ \bibinfo
  {pages} {2718} (\bibinfo {year} {1980})}\BibitemShut {NoStop}%
\bibitem [{\citenamefont {Workman}\ \emph {et~al.}(2022)\citenamefont {Workman}
  \emph {et~al.}}]{pdg2022}%
  \BibitemOpen
  \bibfield  {author} {\bibinfo {author} {\bibfnamefont {R.~L.}\ \bibnamefont
  {Workman}} \emph {et~al.} (\bibinfo {collaboration} {Particle Data Group}),\
  }\href {https://doi.org/10.1093/ptep/ptac097} {\bibfield  {journal} {\bibinfo
   {journal} {PTEP}\ }\textbf {\bibinfo {volume} {2022}},\ \bibinfo {pages}
  {083C01} (\bibinfo {year} {2022})}\BibitemShut {NoStop}%
\bibitem [{\citenamefont {Fukuda}\ \emph {et~al.}(2003)\citenamefont {Fukuda}
  \emph {et~al.}}]{skdet2003}%
  \BibitemOpen
  \bibfield  {author} {\bibinfo {author} {\bibfnamefont {S.}~\bibnamefont
  {Fukuda}} \emph {et~al.},\ }\href
  {https://doi.org/https://doi.org/10.1016/S0168-9002(03)00425-X} {\bibfield
  {journal} {\bibinfo  {journal} {Nucl. Instrum. Methods A}\ }\textbf {\bibinfo
  {volume} {501}},\ \bibinfo {pages} {418} (\bibinfo {year}
  {2003})}\BibitemShut {NoStop}%
\bibitem [{\citenamefont {Abe}\ \emph {et~al.}(2014)\citenamefont {Abe} \emph
  {et~al.}}]{skdetector_2014}%
  \BibitemOpen
  \bibfield  {author} {\bibinfo {author} {\bibfnamefont {K.}~\bibnamefont
  {Abe}} \emph {et~al.} (\bibinfo {collaboration} {Super-Kamiokande
  Collaboration}),\ }\href {https://doi.org/10.1016/j.nima.2013.11.081}
  {\bibfield  {journal} {\bibinfo  {journal} {Nucl. Instrum. Methods A}\
  }\textbf {\bibinfo {volume} {737}},\ \bibinfo {pages} {253–272} (\bibinfo
  {year} {2014})}\BibitemShut {NoStop}%
\bibitem [{\citenamefont {Nishino}\ \emph {et~al.}(2009)\citenamefont {Nishino}
  \emph {et~al.}}]{sk_nim_2009}%
  \BibitemOpen
  \bibfield  {author} {\bibinfo {author} {\bibfnamefont {H.}~\bibnamefont
  {Nishino}} \emph {et~al.},\ }\href
  {https://doi.org/https://doi.org/10.1016/j.nima.2009.09.026} {\bibfield
  {journal} {\bibinfo  {journal} {Nucl. Instrum. Methods A}\ }\textbf {\bibinfo
  {volume} {610}},\ \bibinfo {pages} {710} (\bibinfo {year}
  {2009})}\BibitemShut {NoStop}%
\bibitem [{\citenamefont {Yamada}\ \emph {et~al.}(2010)\citenamefont {Yamada}
  \emph {et~al.}}]{sk_ieee_2010}%
  \BibitemOpen
  \bibfield  {author} {\bibinfo {author} {\bibfnamefont {S.}~\bibnamefont
  {Yamada}} \emph {et~al.},\ }\href {https://doi.org/10.1109/TNS.2009.2034854}
  {\bibfield  {journal} {\bibinfo  {journal} {IEEE Trans. Nucl. Sci.}\ }\textbf
  {\bibinfo {volume} {57}},\ \bibinfo {pages} {428} (\bibinfo {year}
  {2010})}\BibitemShut {NoStop}%
\bibitem [{\citenamefont {Abe}\ \emph {et~al.}(2022{\natexlab{a}})\citenamefont
  {Abe} \emph {et~al.}}]{first_gd_loading}%
  \BibitemOpen
  \bibfield  {author} {\bibinfo {author} {\bibfnamefont {K.}~\bibnamefont
  {Abe}} \emph {et~al.} (\bibinfo {collaboration} {Super-Kamiokande
  Collaboration}),\ }\href {https://doi.org/10.1016/j.nima.2021.166248}
  {\bibfield  {journal} {\bibinfo  {journal} {Nucl. Instrum. Methods A}\
  }\textbf {\bibinfo {volume} {1027}},\ \bibinfo {pages} {166248} (\bibinfo
  {year} {2022}{\natexlab{a}})}\BibitemShut {NoStop}%
\bibitem [{\citenamefont {Shiozawa}(1999)}]{shiozawa_1999}%
  \BibitemOpen
  \bibfield  {author} {\bibinfo {author} {\bibfnamefont {M.}~\bibnamefont
  {Shiozawa}},\ }\href
  {https://doi.org/https://doi.org/10.1016/S0168-9002(99)00359-9} {\bibfield
  {journal} {\bibinfo  {journal} {Nucl. Instrum. Methods A}\ }\textbf {\bibinfo
  {volume} {433}},\ \bibinfo {pages} {240} (\bibinfo {year}
  {1999})}\BibitemShut {NoStop}%
\bibitem [{\citenamefont {Abe}\ \emph {et~al.}(2022{\natexlab{b}})\citenamefont
  {Abe} \emph {et~al.}}]{neutron2013}%
  \BibitemOpen
  \bibfield  {author} {\bibinfo {author} {\bibfnamefont {K.}~\bibnamefont
  {Abe}} \emph {et~al.} (\bibinfo {collaboration} {Super-Kamiokande
  Collaboration}),\ }\href {https://doi.org/10.1088/1748-0221/17/10/P10029}
  {\bibfield  {journal} {\bibinfo  {journal} {JINST}\ }\textbf {\bibinfo
  {volume} {17}}\bibinfo  {number} { (10)},\ \bibinfo {pages}
  {P10029}}\BibitemShut {NoStop}%
\bibitem [{\citenamefont {Men{\'{e}}ndez}(2017)}]{pablo_2017}%
  \BibitemOpen
\bibfield  {number} {  }\bibfield  {author} {\bibinfo {author} {\bibfnamefont
  {P.~F.}\ \bibnamefont {Men{\'{e}}ndez}},\ }\emph {\bibinfo {title} {Neutrino
  Physics in Present and Future Kamioka Water-C{\v{e}}erenkov Detectors with
  Neutron Tagging}},\ \href@noop {} {Ph.D. thesis},\ \bibinfo  {school}
  {Autonomous University of Madrid} (\bibinfo {year} {2017})\BibitemShut
  {NoStop}%
\bibitem [{\citenamefont {Wendell}\ \emph {et~al.}(2010)\citenamefont {Wendell}
  \emph {et~al.}}]{wendell_2010}%
  \BibitemOpen
  \bibfield  {author} {\bibinfo {author} {\bibfnamefont {R.}~\bibnamefont
  {Wendell}} \emph {et~al.} (\bibinfo {collaboration} {Super-Kamiokande
  Collaboration}),\ }\bibfield  {journal} {\bibinfo  {journal} {Phys. Rev. D}\
  }\textbf {\bibinfo {volume} {81}},\ \href
  {https://doi.org/10.1103/physrevd.81.092004} {10.1103/physrevd.81.092004}
  (\bibinfo {year} {2010})\BibitemShut {NoStop}%
\bibitem [{\citenamefont {Matsumoto}(2020)}]{matsumoto_2020}%
  \BibitemOpen
  \bibfield  {author} {\bibinfo {author} {\bibfnamefont {R.}~\bibnamefont
  {Matsumoto}},\ }\emph {\bibinfo {title} {Development of an atmospheric event
  classification method for {Super}-{Kamiokande}}},\ \href@noop {} {Master's
  thesis},\ \bibinfo  {school} {Tokyo University of Science} (\bibinfo {year}
  {2020})\BibitemShut {NoStop}%
\bibitem [{\citenamefont {Takenaka}\ \emph {et~al.}(2020)\citenamefont
  {Takenaka} \emph {et~al.}}]{takenaka_2020}%
  \BibitemOpen
  \bibfield  {author} {\bibinfo {author} {\bibfnamefont {A.}~\bibnamefont
  {Takenaka}} \emph {et~al.} (\bibinfo {collaboration} {Super-Kamiokande
  Collaboration}),\ }\bibfield  {journal} {\bibinfo  {journal} {Phys. Rev. D}\
  }\textbf {\bibinfo {volume} {102}},\ \href
  {https://doi.org/10.1103/physrevd.102.112011} {10.1103/physrevd.102.112011}
  (\bibinfo {year} {2020})\BibitemShut {NoStop}%
\bibitem [{\citenamefont {Honda}\ \emph {et~al.}(2011)\citenamefont {Honda},
  \citenamefont {Kajita}, \citenamefont {Kasahara},\ and\ \citenamefont
  {Midorikawa}}]{honda_2011}%
  \BibitemOpen
  \bibfield  {author} {\bibinfo {author} {\bibfnamefont {M.}~\bibnamefont
  {Honda}}, \bibinfo {author} {\bibfnamefont {T.}~\bibnamefont {Kajita}},
  \bibinfo {author} {\bibfnamefont {K.}~\bibnamefont {Kasahara}},\ and\
  \bibinfo {author} {\bibfnamefont {S.}~\bibnamefont {Midorikawa}},\ }\bibfield
   {journal} {\bibinfo  {journal} {Phys. Rev. D}\ }\textbf {\bibinfo {volume}
  {83}},\ \href {https://doi.org/10.1103/physrevd.83.123001}
  {10.1103/physrevd.83.123001} (\bibinfo {year} {2011})\BibitemShut {NoStop}%
\bibitem [{\citenamefont {Hayato}(2002)}]{neut}%
  \BibitemOpen
  \bibfield  {author} {\bibinfo {author} {\bibfnamefont {Y.}~\bibnamefont
  {Hayato}},\ }\href {https://doi.org/10.1016/s0920-5632(02)01759-0} {\bibfield
   {journal} {\bibinfo  {journal} {Nucl. Phys. B Proc. Suppl.}\ }\textbf
  {\bibinfo {volume} {112}},\ \bibinfo {pages} {171–176} (\bibinfo {year}
  {2002})}\BibitemShut {NoStop}%
\bibitem [{\citenamefont {Hayato}\ and\ \citenamefont
  {Pickering}(2021)}]{neut2021}%
  \BibitemOpen
  \bibfield  {author} {\bibinfo {author} {\bibfnamefont {Y.}~\bibnamefont
  {Hayato}}\ and\ \bibinfo {author} {\bibfnamefont {L.}~\bibnamefont
  {Pickering}},\ }\href {https://doi.org/10.1140/epjs/s11734-021-00287-7}
  {\bibfield  {journal} {\bibinfo  {journal} {Eur. Phys. J. Spec. Top.}\
  }\textbf {\bibinfo {volume} {230}},\ \bibinfo {pages} {4469} (\bibinfo {year}
  {2021})}\BibitemShut {NoStop}%
\bibitem [{\citenamefont {Brun}\ \emph {et~al.}(1994)\citenamefont {Brun} \emph
  {et~al.}}]{geant_1994}%
  \BibitemOpen
  \bibfield  {author} {\bibinfo {author} {\bibfnamefont {R.}~\bibnamefont
  {Brun}} \emph {et~al.},\ }\href {https://cds.cern.ch/record/1082634} {\emph
  {\bibinfo {title} {{GEANT}: {D}etector Description and Simulation Tool}}},\
  CERN Program Library\ (\bibinfo {year} {1994})\BibitemShut {NoStop}%
\bibitem [{\citenamefont {Smith}\ and\ \citenamefont
  {Moniz}(1972)}]{smith_moniz_1972}%
  \BibitemOpen
  \bibfield  {author} {\bibinfo {author} {\bibfnamefont {R.}~\bibnamefont
  {Smith}}\ and\ \bibinfo {author} {\bibfnamefont {E.}~\bibnamefont {Moniz}},\
  }\href {https://doi.org/10.1016/0550-3213(72)90040-5} {\bibfield  {journal}
  {\bibinfo  {journal} {Nucl. Phys. B}\ }\textbf {\bibinfo {volume} {43}},\
  \bibinfo {pages} {605–622} (\bibinfo {year} {1972})}\BibitemShut {NoStop}%
\bibitem [{\citenamefont {Nieves}\ \emph {et~al.}(2011)\citenamefont {Nieves},
  \citenamefont {Simo},\ and\ \citenamefont {Vacas}}]{nieves2011}%
  \BibitemOpen
  \bibfield  {author} {\bibinfo {author} {\bibfnamefont {J.}~\bibnamefont
  {Nieves}}, \bibinfo {author} {\bibfnamefont {I.~R.}\ \bibnamefont {Simo}},\
  and\ \bibinfo {author} {\bibfnamefont {M.~J.~V.}\ \bibnamefont {Vacas}},\
  }\href {https://doi.org/10.1103/PhysRevC.83.045501} {\bibfield  {journal}
  {\bibinfo  {journal} {Phys. Rev. C}\ }\textbf {\bibinfo {volume} {83}},\
  \bibinfo {pages} {045501} (\bibinfo {year} {2011})}\BibitemShut {NoStop}%
\bibitem [{\citenamefont {Nieves}\ \emph {et~al.}(2016)\citenamefont {Nieves}
  \emph {et~al.}}]{nieves_rpa}%
  \BibitemOpen
  \bibfield  {author} {\bibinfo {author} {\bibfnamefont {J.}~\bibnamefont
  {Nieves}} \emph {et~al.},\ }\href
  {https://doi.org/10.1016/j.nuclphysbps.2015.09.295} {\bibfield  {journal}
  {\bibinfo  {journal} {Nucl. Part. Phys. Proc.}\ }\textbf {\bibinfo {volume}
  {273-275}},\ \bibinfo {pages} {1830–1835} (\bibinfo {year}
  {2016})}\BibitemShut {NoStop}%
\bibitem [{\citenamefont {Bradford}\ \emph {et~al.}(2006)\citenamefont
  {Bradford}, \citenamefont {Bodek}, \citenamefont {Budd},\ and\ \citenamefont
  {Arrington}}]{bbba05}%
  \BibitemOpen
  \bibfield  {author} {\bibinfo {author} {\bibfnamefont {R.}~\bibnamefont
  {Bradford}}, \bibinfo {author} {\bibfnamefont {A.}~\bibnamefont {Bodek}},
  \bibinfo {author} {\bibfnamefont {H.}~\bibnamefont {Budd}},\ and\ \bibinfo
  {author} {\bibfnamefont {J.}~\bibnamefont {Arrington}},\ }\href
  {https://doi.org/https://doi.org/10.1016/j.nuclphysbps.2006.08.028}
  {\bibfield  {journal} {\bibinfo  {journal} {Nucl. Phys. B Proc. Suppl.}\
  }\textbf {\bibinfo {volume} {159}},\ \bibinfo {pages} {127} (\bibinfo {year}
  {2006})}\BibitemShut {NoStop}%
\bibitem [{\citenamefont {Rein}\ and\ \citenamefont
  {Sehgal}(1981)}]{rein_sehgal_1981}%
  \BibitemOpen
  \bibfield  {author} {\bibinfo {author} {\bibfnamefont {D.}~\bibnamefont
  {Rein}}\ and\ \bibinfo {author} {\bibfnamefont {L.~M.}\ \bibnamefont
  {Sehgal}},\ }\href {https://doi.org/10.1016/0003-4916(81)90242-6} {\bibfield
  {journal} {\bibinfo  {journal} {Ann. Phys.}\ }\textbf {\bibinfo {volume}
  {133}},\ \bibinfo {pages} {79–153} (\bibinfo {year} {1981})}\BibitemShut
  {NoStop}%
\bibitem [{\citenamefont {Berger}\ and\ \citenamefont
  {Sehgal}(2007)}]{berger_sehgal_2007}%
  \BibitemOpen
  \bibfield  {author} {\bibinfo {author} {\bibfnamefont {C.}~\bibnamefont
  {Berger}}\ and\ \bibinfo {author} {\bibfnamefont {L.~M.}\ \bibnamefont
  {Sehgal}},\ }\bibfield  {journal} {\bibinfo  {journal} {Phys. Rev. D}\
  }\textbf {\bibinfo {volume} {76}},\ \href
  {https://doi.org/10.1103/physrevd.76.113004} {10.1103/physrevd.76.113004}
  (\bibinfo {year} {2007})\BibitemShut {NoStop}%
\bibitem [{\citenamefont {Martins}(2016)}]{martins2016charged}%
  \BibitemOpen
  \bibfield  {author} {\bibinfo {author} {\bibfnamefont {P.}~\bibnamefont
  {Martins}},\ }\href@noop {} {\bibinfo {title} {Charged current coherent pion
  production in neutrino scattering}} (\bibinfo {year} {2016}),\ \Eprint
  {https://arxiv.org/abs/1605.00095} {arXiv:1605.00095 [hep-ex]} \BibitemShut
  {NoStop}%
\bibitem [{\citenamefont {Glück}\ \emph {et~al.}(1998)\citenamefont {Glück},
  \citenamefont {Reya},\ and\ \citenamefont {Vogt}}]{grv98}%
  \BibitemOpen
  \bibfield  {author} {\bibinfo {author} {\bibfnamefont {M.}~\bibnamefont
  {Glück}}, \bibinfo {author} {\bibfnamefont {E.}~\bibnamefont {Reya}},\ and\
  \bibinfo {author} {\bibfnamefont {A.}~\bibnamefont {Vogt}},\ }\href
  {https://doi.org/10.1007/s100529800978} {\bibfield  {journal} {\bibinfo
  {journal} {Eur. Phys. J. C}\ }\textbf {\bibinfo {volume} {5}},\ \bibinfo
  {pages} {461–470} (\bibinfo {year} {1998})}\BibitemShut {NoStop}%
\bibitem [{\citenamefont {Bodek}\ and\ \citenamefont
  {Yang}(2005)}]{bodek_yang}%
  \BibitemOpen
  \bibfield  {author} {\bibinfo {author} {\bibfnamefont {A.}~\bibnamefont
  {Bodek}}\ and\ \bibinfo {author} {\bibfnamefont {U.}~\bibnamefont {Yang}},\
  }\href {https://doi.org/10.1063/1.2122031} {\bibfield  {journal} {\bibinfo
  {journal} {AIP Conf. Proc.}\ }\textbf {\bibinfo {volume} {792}},\ \bibinfo
  {pages} {257–260} (\bibinfo {year} {2005})}\BibitemShut {NoStop}%
\bibitem [{\citenamefont {Sj{\"o}strand}(1994)}]{pythia572}%
  \BibitemOpen
  \bibfield  {author} {\bibinfo {author} {\bibfnamefont {T.}~\bibnamefont
  {Sj{\"o}strand}},\ }\href@noop {} {\bibfield  {journal} {\bibinfo  {journal}
  {Comput. Phys. Commun.}\ }\textbf {\bibinfo {volume} {82}},\ \bibinfo {pages}
  {74} (\bibinfo {year} {1994})}\BibitemShut {NoStop}%
\bibitem [{\citenamefont {Pinzon~Guerra}\ \emph {et~al.}(2017)\citenamefont
  {Pinzon~Guerra} \emph {et~al.}}]{duet2017}%
  \BibitemOpen
  \bibfield  {author} {\bibinfo {author} {\bibfnamefont {E.~S.}\ \bibnamefont
  {Pinzon~Guerra}} \emph {et~al.} (\bibinfo {collaboration} {DUET
  Collaboration}),\ }\href {https://doi.org/10.1103/PhysRevC.95.045203}
  {\bibfield  {journal} {\bibinfo  {journal} {Phys. Rev. C}\ }\textbf {\bibinfo
  {volume} {95}},\ \bibinfo {pages} {045203} (\bibinfo {year}
  {2017})}\BibitemShut {NoStop}%
\bibitem [{\citenamefont {Gran}\ \emph {et~al.}(2013)\citenamefont {Gran},
  \citenamefont {Nieves}, \citenamefont {Sanchez},\ and\ \citenamefont
  {Vacas}}]{nieves_2p2h_2013}%
  \BibitemOpen
  \bibfield  {author} {\bibinfo {author} {\bibfnamefont {R.}~\bibnamefont
  {Gran}}, \bibinfo {author} {\bibfnamefont {J.}~\bibnamefont {Nieves}},
  \bibinfo {author} {\bibfnamefont {F.}~\bibnamefont {Sanchez}},\ and\ \bibinfo
  {author} {\bibfnamefont {M.~J.}\ \bibnamefont {Vacas}},\ }\bibfield
  {journal} {\bibinfo  {journal} {Phys. Rev. D}\ }\textbf {\bibinfo {volume}
  {88}},\ \href {https://doi.org/10.1103/physrevd.88.113007}
  {10.1103/physrevd.88.113007} (\bibinfo {year} {2013})\BibitemShut {NoStop}%
\bibitem [{\citenamefont {Wester}(2023)}]{twester_2023}%
  \BibitemOpen
  \bibfield  {author} {\bibinfo {author} {\bibfnamefont {T.}~\bibnamefont
  {Wester}},\ }\emph {\bibinfo {title} {Discerning the Neutrino Mass Ordering
  using Atmospheric Neutrinos in Super-Kamiokande I--V}},\ \href@noop {} {Ph.D.
  thesis},\ \bibinfo  {school} {Boston University} (\bibinfo {year}
  {2023})\BibitemShut {NoStop}%
\bibitem [{\citenamefont {Honda}\ \emph {et~al.}(2007)\citenamefont {Honda},
  \citenamefont {Kajita}, \citenamefont {Kasahara}, \citenamefont
  {Midorikawa},\ and\ \citenamefont {Sanuki}}]{honda_2007}%
  \BibitemOpen
  \bibfield  {author} {\bibinfo {author} {\bibfnamefont {M.}~\bibnamefont
  {Honda}}, \bibinfo {author} {\bibfnamefont {T.}~\bibnamefont {Kajita}},
  \bibinfo {author} {\bibfnamefont {K.}~\bibnamefont {Kasahara}}, \bibinfo
  {author} {\bibfnamefont {S.}~\bibnamefont {Midorikawa}},\ and\ \bibinfo
  {author} {\bibfnamefont {T.}~\bibnamefont {Sanuki}},\ }\href
  {https://doi.org/10.1103/PhysRevD.75.043006} {\bibfield  {journal} {\bibinfo
  {journal} {Phys. Rev. D}\ }\textbf {\bibinfo {volume} {75}},\ \bibinfo
  {pages} {043006} (\bibinfo {year} {2007})}\BibitemShut {NoStop}%
\bibitem [{\citenamefont {Abe}\ \emph {et~al.}(2015)\citenamefont {Abe} \emph
  {et~al.}}]{t2k_2015}%
  \BibitemOpen
  \bibfield  {author} {\bibinfo {author} {\bibfnamefont {K.}~\bibnamefont
  {Abe}} \emph {et~al.} (\bibinfo {collaboration} {T2K Collaboration}),\ }\href
  {https://doi.org/10.1103/PhysRevD.91.072010} {\bibfield  {journal} {\bibinfo
  {journal} {Phys. Rev. D}\ }\textbf {\bibinfo {volume} {91}},\ \bibinfo
  {pages} {072010} (\bibinfo {year} {2015})}\BibitemShut {NoStop}%
\bibitem [{\citenamefont {Akutsu}(2020)}]{akutsu_2020}%
  \BibitemOpen
  \bibfield  {author} {\bibinfo {author} {\bibfnamefont {R.}~\bibnamefont
  {Akutsu}},\ }\emph {\bibinfo {title} {A Study of Neutrons Associated with
  Neutrino and Antineutrino Interactions on the Water Target at the T2K Far
  Detector}},\ \href@noop {} {Ph.D. thesis},\ \bibinfo  {school} {University of
  Tokyo} (\bibinfo {year} {2020})\BibitemShut {NoStop}%
\bibitem [{\citenamefont {Andreopoulos}\ \emph {et~al.}(2010)\citenamefont
  {Andreopoulos} \emph {et~al.}}]{Andreopoulos:2009rq}%
  \BibitemOpen
  \bibfield  {author} {\bibinfo {author} {\bibfnamefont {C.}~\bibnamefont
  {Andreopoulos}} \emph {et~al.},\ }\href
  {https://doi.org/10.1016/j.nima.2009.12.009} {\bibfield  {journal} {\bibinfo
  {journal} {Nucl. Instrum. Methods A}\ }\textbf {\bibinfo {volume} {614}},\
  \bibinfo {pages} {87} (\bibinfo {year} {2010})},\ \Eprint
  {https://arxiv.org/abs/0905.2517} {arXiv:0905.2517 [hep-ph]} \BibitemShut
  {NoStop}%
\bibitem [{\citenamefont {Abe}\ \emph {et~al.}(2013)\citenamefont {Abe} \emph
  {et~al.}}]{sk_tau_2013}%
  \BibitemOpen
  \bibfield  {author} {\bibinfo {author} {\bibfnamefont {K.}~\bibnamefont
  {Abe}} \emph {et~al.} (\bibinfo {collaboration} {Super-Kamiokande
  Collaboration}),\ }\href {https://doi.org/10.1103/PhysRevLett.110.181802}
  {\bibfield  {journal} {\bibinfo  {journal} {Phys. Rev. Lett.}\ }\textbf
  {\bibinfo {volume} {110}},\ \bibinfo {pages} {181802} (\bibinfo {year}
  {2013})}\BibitemShut {NoStop}%
\bibitem [{\citenamefont {Hagiwara}\ \emph {et~al.}(2003)\citenamefont
  {Hagiwara}, \citenamefont {Mawatari},\ and\ \citenamefont
  {Yokoya}}]{hagiwara_2003}%
  \BibitemOpen
  \bibfield  {author} {\bibinfo {author} {\bibfnamefont {K.}~\bibnamefont
  {Hagiwara}}, \bibinfo {author} {\bibfnamefont {K.}~\bibnamefont {Mawatari}},\
  and\ \bibinfo {author} {\bibfnamefont {H.}~\bibnamefont {Yokoya}},\ }\href
  {https://doi.org/https://doi.org/10.1016/S0550-3213(03)00575-3} {\bibfield
  {journal} {\bibinfo  {journal} {Nucl. Phys. B}\ }\textbf {\bibinfo {volume}
  {668}},\ \bibinfo {pages} {364} (\bibinfo {year} {2003})}\BibitemShut
  {NoStop}%
\bibitem [{\citenamefont {Watanabe}\ \emph {et~al.}(2009)\citenamefont
  {Watanabe} \emph {et~al.}}]{first_ntag_2009}%
  \BibitemOpen
  \bibfield  {author} {\bibinfo {author} {\bibfnamefont {H.}~\bibnamefont
  {Watanabe}} \emph {et~al.},\ }\href
  {https://doi.org/https://doi.org/10.1016/j.astropartphys.2009.03.002}
  {\bibfield  {journal} {\bibinfo  {journal} {Astropart. Phys.}\ }\textbf
  {\bibinfo {volume} {31}},\ \bibinfo {pages} {320} (\bibinfo {year}
  {2009})}\BibitemShut {NoStop}%
\bibitem [{\citenamefont {Li}\ \emph {et~al.}(2018)\citenamefont {Li} \emph
  {et~al.}}]{li2018}%
  \BibitemOpen
  \bibfield  {author} {\bibinfo {author} {\bibfnamefont {Z.}~\bibnamefont {Li}}
  \emph {et~al.} (\bibinfo {collaboration} {Super-Kamiokande Collaboration}),\
  }\href {https://doi.org/10.1103/PhysRevD.98.052006} {\bibfield  {journal}
  {\bibinfo  {journal} {Phys. Rev. D}\ }\textbf {\bibinfo {volume} {98}},\
  \bibinfo {pages} {052006} (\bibinfo {year} {2018})}\BibitemShut {NoStop}%
\bibitem [{\citenamefont {Adamson}\ \emph {et~al.}(2020)\citenamefont {Adamson}
  \emph {et~al.}}]{minos_2020}%
  \BibitemOpen
  \bibfield  {author} {\bibinfo {author} {\bibfnamefont {P.}~\bibnamefont
  {Adamson}} \emph {et~al.} (\bibinfo {collaboration} {$\mathrm{MINOS}+$
  Collaboration}),\ }\href {https://doi.org/10.1103/PhysRevLett.125.131802}
  {\bibfield  {journal} {\bibinfo  {journal} {Phys. Rev. Lett.}\ }\textbf
  {\bibinfo {volume} {125}},\ \bibinfo {pages} {131802} (\bibinfo {year}
  {2020})}\BibitemShut {NoStop}%
\bibitem [{\citenamefont {Abe}\ \emph {et~al.}(2023)\citenamefont {Abe} \emph
  {et~al.}}]{t2k_2023}%
  \BibitemOpen
  \bibfield  {author} {\bibinfo {author} {\bibfnamefont {K.}~\bibnamefont
  {Abe}} \emph {et~al.} (\bibinfo {collaboration} {T2K Collaboration}),\ }\href
  {https://doi.org/10.1140/epjc/s10052-023-11819-x} {\bibfield  {journal}
  {\bibinfo  {journal} {Eur. Phys. J. C}\ }\textbf {\bibinfo {volume} {83}},\
  \bibinfo {pages} {782} (\bibinfo {year} {2023})}\BibitemShut {NoStop}%
\bibitem [{\citenamefont {Abbasi}\ \emph {et~al.}(2023)\citenamefont {Abbasi}
  \emph {et~al.}}]{icecube_2023}%
  \BibitemOpen
  \bibfield  {author} {\bibinfo {author} {\bibfnamefont {R.}~\bibnamefont
  {Abbasi}} \emph {et~al.} (\bibinfo {collaboration} {IceCube Collaboration}),\
  }\href {https://doi.org/10.1103/PhysRevD.108.012014} {\bibfield  {journal}
  {\bibinfo  {journal} {Phys. Rev. D}\ }\textbf {\bibinfo {volume} {108}},\
  \bibinfo {pages} {012014} (\bibinfo {year} {2023})}\BibitemShut {NoStop}%
\bibitem [{\citenamefont {Himmel}(2020)}]{alex_himmel_2020_4142045}%
  \BibitemOpen
  \bibfield  {author} {\bibinfo {author} {\bibfnamefont {A.}~\bibnamefont
  {Himmel}},\ }\href {https://doi.org/10.5281/zenodo.4142045} {\bibinfo {title}
  {New oscillation results from the {NOvA} experiment}} (\bibinfo {year}
  {2020}),\ \bibinfo {note} {10.5281/zenodo.4142045}\BibitemShut {NoStop}%
\bibitem [{\citenamefont {Algeri}\ \emph {et~al.}(2020)\citenamefont {Algeri},
  \citenamefont {Aalbers}, \citenamefont {Mor{\aa}},\ and\ \citenamefont
  {Conrad}}]{algeri2020}%
  \BibitemOpen
  \bibfield  {author} {\bibinfo {author} {\bibfnamefont {S.}~\bibnamefont
  {Algeri}}, \bibinfo {author} {\bibfnamefont {J.}~\bibnamefont {Aalbers}},
  \bibinfo {author} {\bibfnamefont {K.~D.}\ \bibnamefont {Mor{\aa}}},\ and\
  \bibinfo {author} {\bibfnamefont {J.}~\bibnamefont {Conrad}},\ }\href
  {https://doi.org/10.1038/s42254-020-0169-5} {\bibfield  {journal} {\bibinfo
  {journal} {Nature Reviews Physics}\ }\textbf {\bibinfo {volume} {2}},\
  \bibinfo {pages} {245} (\bibinfo {year} {2020})}\BibitemShut {NoStop}%
\bibitem [{\citenamefont {Read}(2002)}]{cls_2002}%
  \BibitemOpen
  \bibfield  {author} {\bibinfo {author} {\bibfnamefont {A.~L.}\ \bibnamefont
  {Read}},\ }\href {https://doi.org/10.1088/0954-3899/28/10/313} {\bibfield
  {journal} {\bibinfo  {journal} {J. Phys. G}\ }\textbf {\bibinfo {volume}
  {28}},\ \bibinfo {pages} {2693–2704} (\bibinfo {year} {2002})}\BibitemShut
  {NoStop}%
\bibitem [{sk_(2023)}]{sk_data_release_2023}%
  \BibitemOpen
  \href {https://doi.org/10.5281/zenodo.8401262} {\bibinfo {title} {Data
  release: Atmospheric neutrino oscillation analysis with neutron tagging and
  an expanded fiducial volume in {S}uper-{K}amiokande {I}-{V}}} (\bibinfo
  {year} {2023}),\ \bibinfo {note} {10.5281/zenodo.8401262}\BibitemShut
  {NoStop}%
\end{thebibliography}%

\end{document}